\documentclass[12pt]{article}
\input epsf
\usepackage{epsfig}
\usepackage{latexsym}
\usepackage{amsfonts}
\newtheorem{theorem}{Theorem}
\newtheorem{remark}{Remark}
\newtheorem{proposition}{Proposition}

\def\C{{\mathbf{C}}}
\def\bC{{\mathbf{\overline{\mathbf{C}}}}}

\begin{document}
\title{Irreducibility of some spectral determinants}
\author{Alexandre Eremenko\thanks{Supported by NSF grant
DMS-0555279.}{$\;$} and Andrei Gabrielov\thanks{supported by NSF grant DMS-0801050.}}
\maketitle
\begin{abstract}
This is a complement to our paper arXiv:0802.1461.
We study irreducibility of spectral determinants
of some one-parametric eigenvalue problems in dimension one
with polynomial potentials.
\end{abstract}

We consider eigenvalue problems of the type
$$-y^{\prime\prime}+P(\alpha,z)y=\lambda y,$$
with the boundary condition that $y(z)\to 0$
along appropriately chosen paths in the complex plane
(for example, at $\pm\infty$ on the real line). The potential $P$
is a polynomial in the independent variable $z$ depending
on a complex parameter $\alpha$. The problems we consider
have discrete spectra for all complex $\alpha$.
The dependence of the eigenvalues on $\alpha$ is described
by the equation
$$F(\alpha,\lambda)=0,$$
where $F$ is an entire function of two variables which is
called the spectral determinant. We study irreducibility
of the spectral determinant for certain eigenvalue problems
that occur in quantum mechanics.

Problems of this type were considered for the first time
in \cite{Schafke1,Schafke2} for the case of Mathieu equation
with the boundary conditions on a finite interval.
We refer to \cite{Volkmar,Mit1,Mit2,Mit3}
for further development.

Our research was stimulated by the paper of Bender and Wu
who discovered in \cite{BW} that the spectral determinant
of the even quartic oscillator has exactly two irreducible
components. In our paper \cite{EG} we gave a complete
rigorous proof
of this fact. It was mentioned in
\cite{EG} that our method applies to several
other one-parametric families of eigenvalue problems,
and the purpose of the present paper is to give some
details of these applications.

We begin with a sketch of the main result of
\cite{EG} referring to that paper for complete 
details.
\vspace{.2in}

\begin{center}
{\bf Even quartic oscillator}
\end{center}
\vspace{.1in}

1. Eigenvalue problem for the even quartic oscillator:
\begin{equation}
\label{1}
-y^{\prime\prime}+(z^4+\alpha z^2)y=\lambda y,
\end{equation}
\begin{equation}
\label{2}
y(\pm\infty)=0\quad\mbox{on the real line},
\end{equation}
where $\alpha$ is a complex parameter.
\newline
The spectrum is discrete, infinite, simple.
$$Z=\{(\alpha,\lambda)\in\C^2:\lambda\;\mbox{is an eigenvalue
of (1), (2)}\}$$
is an analytic subset of $\C^2$.

\begin{theorem} The set $Z$ consists of two irreducible
components: one for even eigenfunctions, another for
odd ones.
These irreducible components are also connected components.
Moreover, the set $Z$ is non-singular.
\end{theorem}
\vspace{.1in}

2. Parametrization of the set $Z$.

Let $G$ be the set of all odd meromorphic functions $f$
such that $f(z)\to 0$ as $z\to\pm\infty$ on the real line,
and the Schwarzian
$$S_f=\frac{f^{\prime\prime\prime}}{f'}
-\frac{3}{2}\left(\frac{f^{\prime\prime}}{f'}\right)^2$$
is a polynomial of degree $4$ with leading coefficient $-2$:
$$-\frac{1}{2}S_f(z)=z^4+\alpha z^2-\lambda.$$
We have a map $\Phi:G\to\C^2,\; f\mapsto(\alpha,\lambda).$

\begin{proposition} $\Phi$ maps $G$ to $Z$ surjectively.
\end{proposition}

\begin{remark} If we define the equivalence relation by
$f_1\sim f_2$ if $f_1=cf_2$ then $\tilde{\Phi}:G/\!\!\sim\;\to Z$
is a biholomorphic parametrization.
\end{remark}

3. Sketch of the proof. $f=y/y_1$ where $y$ is an eigenfunction
(it is always even or odd), and $y_1$ a linearly independent
solution of {\em opposite parity}. In the opposite direction:
$y_1=1/\sqrt{f'},\; y=fy_1$. 

\begin{eqnarray*}
&f(z)\to 0,\quad |z|\to\infty\quad\mbox{on the real axis}\\
\Longleftrightarrow&y\; \mbox{is subdominant in}\; S_0,S_3\\
\Longleftrightarrow&y\; \mbox{is an eigenfunction}\\
\Longleftrightarrow&\lambda\; \mbox{is an eigenvalue}.
\end{eqnarray*}

\bigskip
\begin{center}
\epsfxsize=2.5in%
\centerline{\epsffile{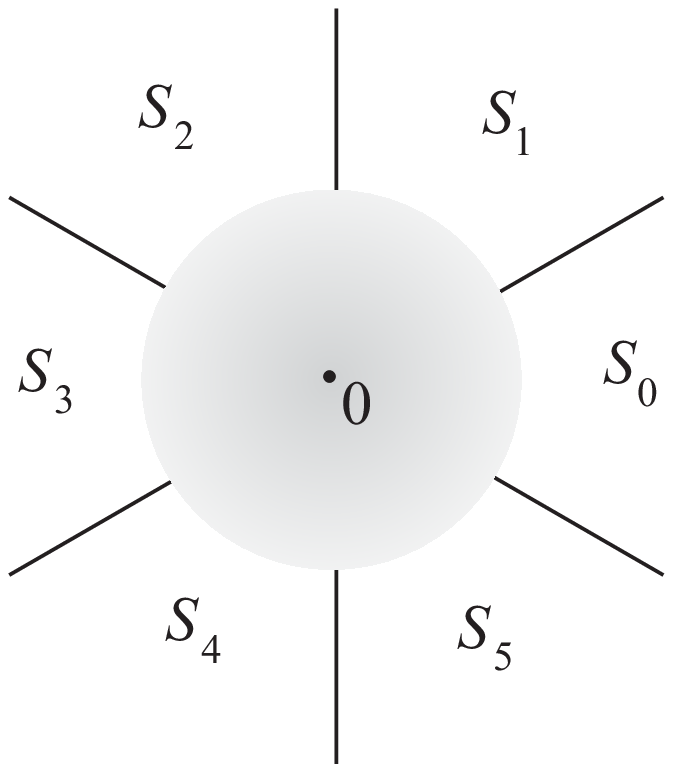}}
\nopagebreak
\noindent Fig 1. Stokes sectors
\end{center}
\bigskip

4. Deformation of functions in $G$.

\medskip
\noindent $f\in G$ has no critical points:
$f'=(y'y_1-yy'_1)/y_1^2$ so $f'(z)\neq 0$ and the poles are
simple.

\smallskip
\noindent $f$ has $6$ {\em asymptotic tracts} corresponding to the Stokes
sectors: $f(z)\to w_j$ in $S_j$.

Thus 
$$f:\C\backslash f^{-1}(\{w_0,\ldots,w_5\})\to\bC\backslash
\{ w_0,\ldots,w_5\}$$
is an unramified covering.
\begin{proposition} Let $f_0\in G$, and let $w_j(t)$ be
a centrally symmetric deformation of the asymptotic values 
such that
$$w_j(0)\neq w_k(0)\Longrightarrow w_j(t)\neq w_k(t),
\quad t\in[0,1].$$ 
Then there exists a deformation $f_t\in G$ for
$t\in[0,1]$ such that $f_t(z)\to w_j(t)$ in $S_j$.
\end{proposition}
\vspace{.1in}

5. Sketch of the proof. Let $\psi_t:\bC\to\bC$ be
odd diffeomorphisms, $\psi_t(w_j(0))=w_j(t)$.
Then there exist odd diffeomorphisms $\phi_t:\C\to\C$
such that $g_t=\psi_t\circ f_0\circ\phi_t$ are
meromorphic functions in $\C$ (Nevanlinna, 1930). These functions
have no critical points and $6$ asymptotic tracts
with asymptotic values $w_j(t)$.

For every meromorphic function with no critical points
and $q$ asymptotic tracts, the
Schwarzian $S_f$ is a polynomial of degree $q-2$ (Nevanlinna, 1930).
Putting $f_t(z)=g(c_tz)$ we make $-(1/2)S_f$ monic;
it is even because $f$ is odd.
\vspace{.1in}

6. We may assume that
$$(w_0,w_1,w_2,w_3,w_4,w_5)=
(0,i,1,0,-i,-1).$$
How to describe all $f\in G$ with such asymptotic values?

Let $\Psi_f=f^{-1}(\Psi_0)$, where $\Psi_0$ 
is the following cell decomposition of the Riemann sphere:
\vspace{.3in}

\begin{center}
\epsfxsize=4.0in%
\centerline{\epsffile{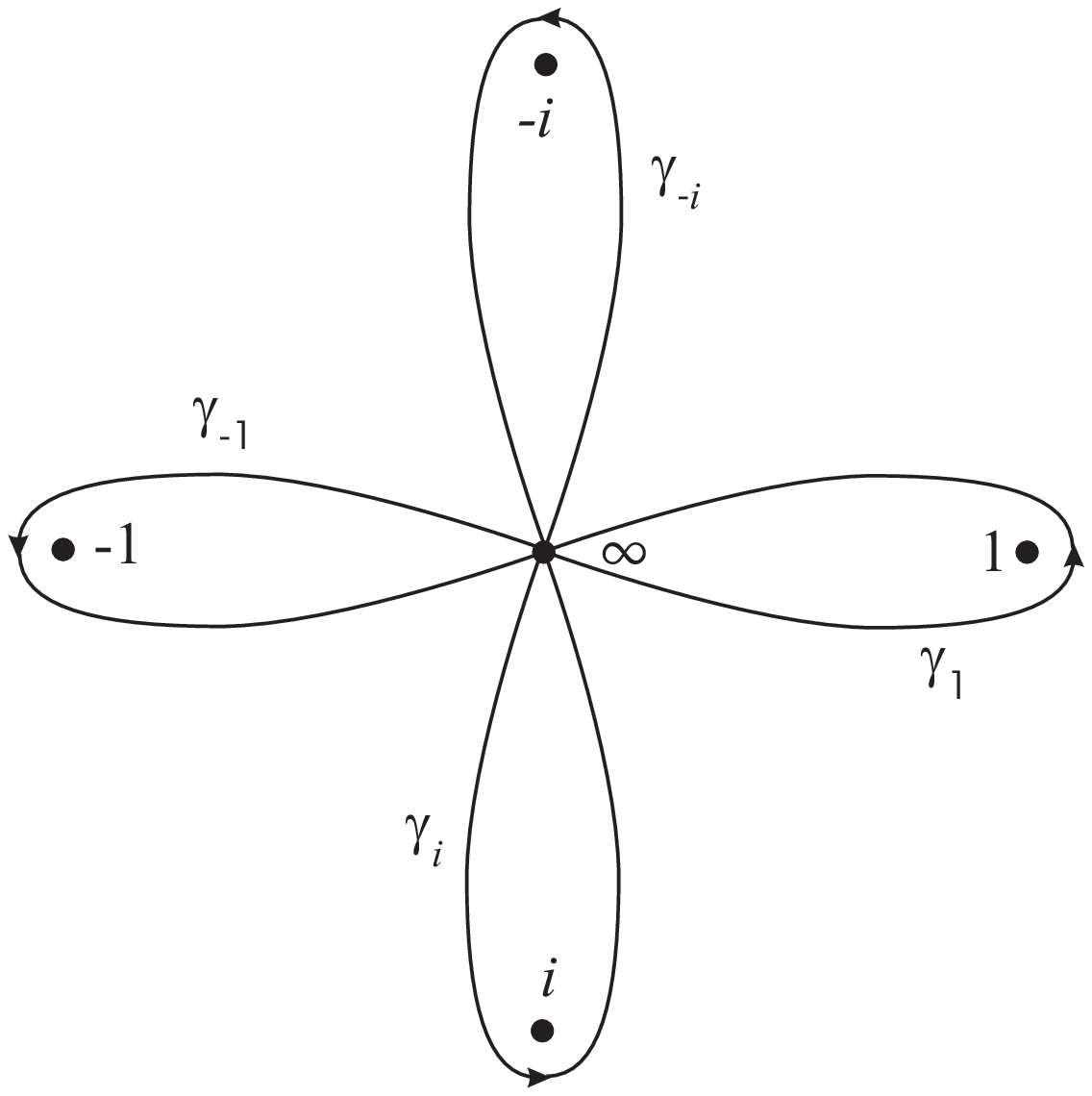}}
\nopagebreak
\vspace{.2in}

\noindent Fig 2. Cell decomposition $\Psi_0$.
\end{center}
The next figure shows how the preimage of $\Psi_0$ may look.
\newpage
\begin{center}
\epsfxsize=4.0in%
\centerline{\epsffile{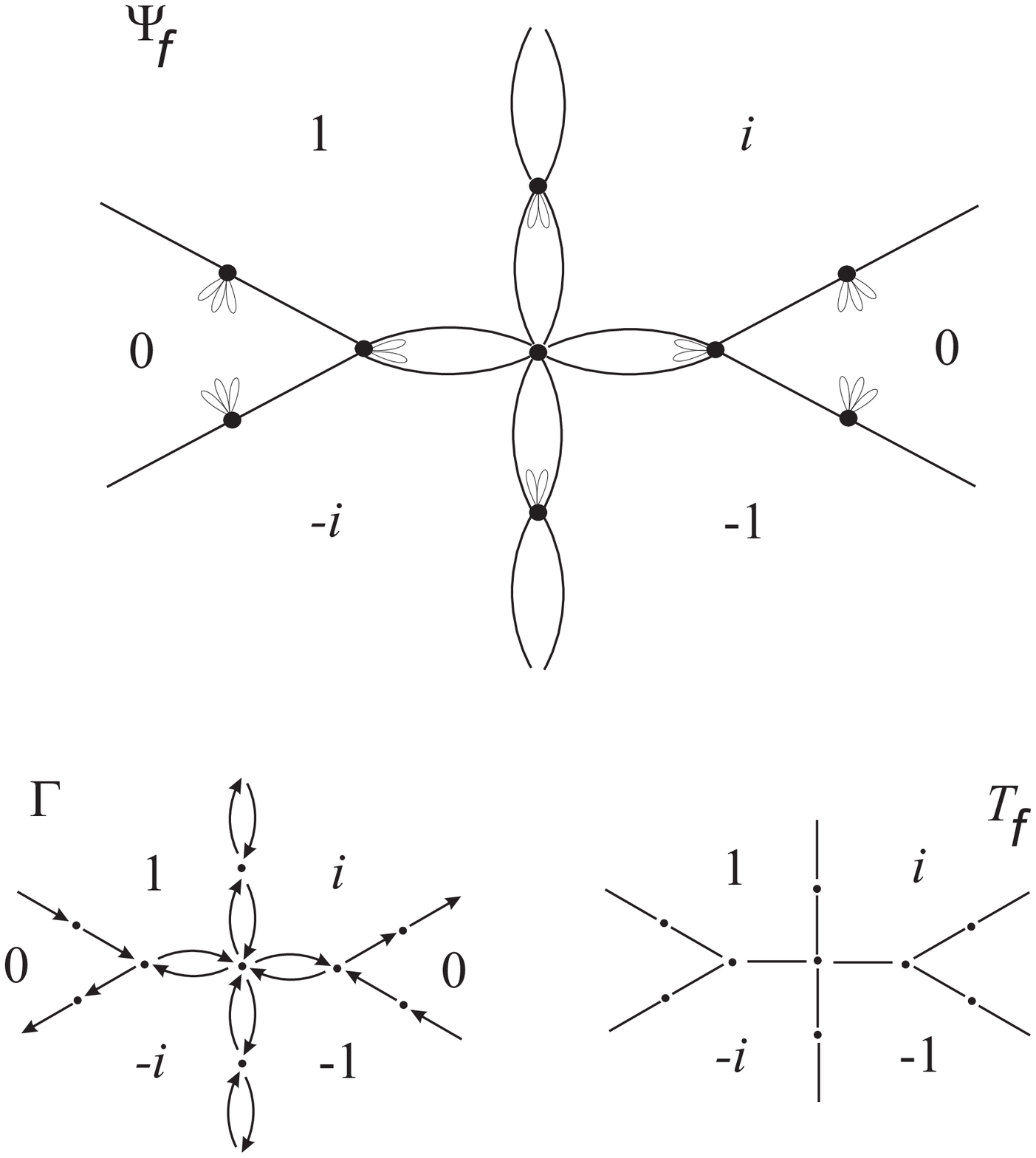}}
\nopagebreak
\vspace{.2in}
\noindent Fig 3. Cell decomposition $\Psi_f$.
\end{center}
\vspace{.2in}

Faces of $\Psi_f$ are labeled by asymptotic values.
Removing loops and replacing multiple edges of $\Psi_f$
by simple edges, we obtain a {\em tree} $T$.
This tree has $6$ faces and faces labeled $0$ cannot
have a common boundary edge. 
It is possible to classify all such trees.
\newpage

\begin{center}
\epsfxsize=6.5in%
\centerline{\epsffile{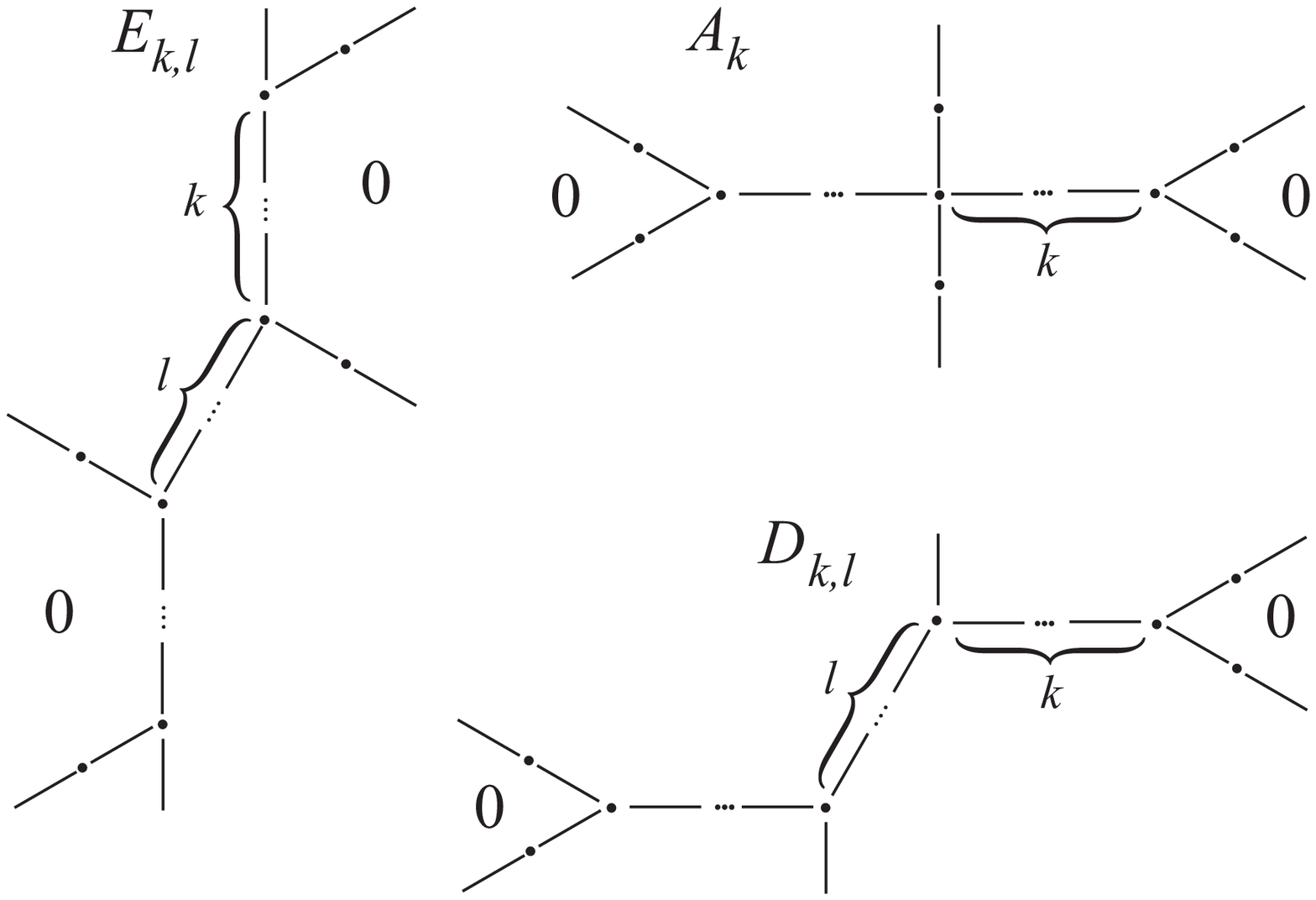}}
\nopagebreak
\vspace{.2in}

\noindent Fig 4. Classification of trees: $A_k:k\geq 0$,
$D_{k,l}:k\geq 0,l\geq 1,$ $E_{k,l}:k\geq 1,l\geq 0$.
\end{center}
\newpage

\begin{center}
\epsfxsize=4.5in%
\centerline{\epsffile{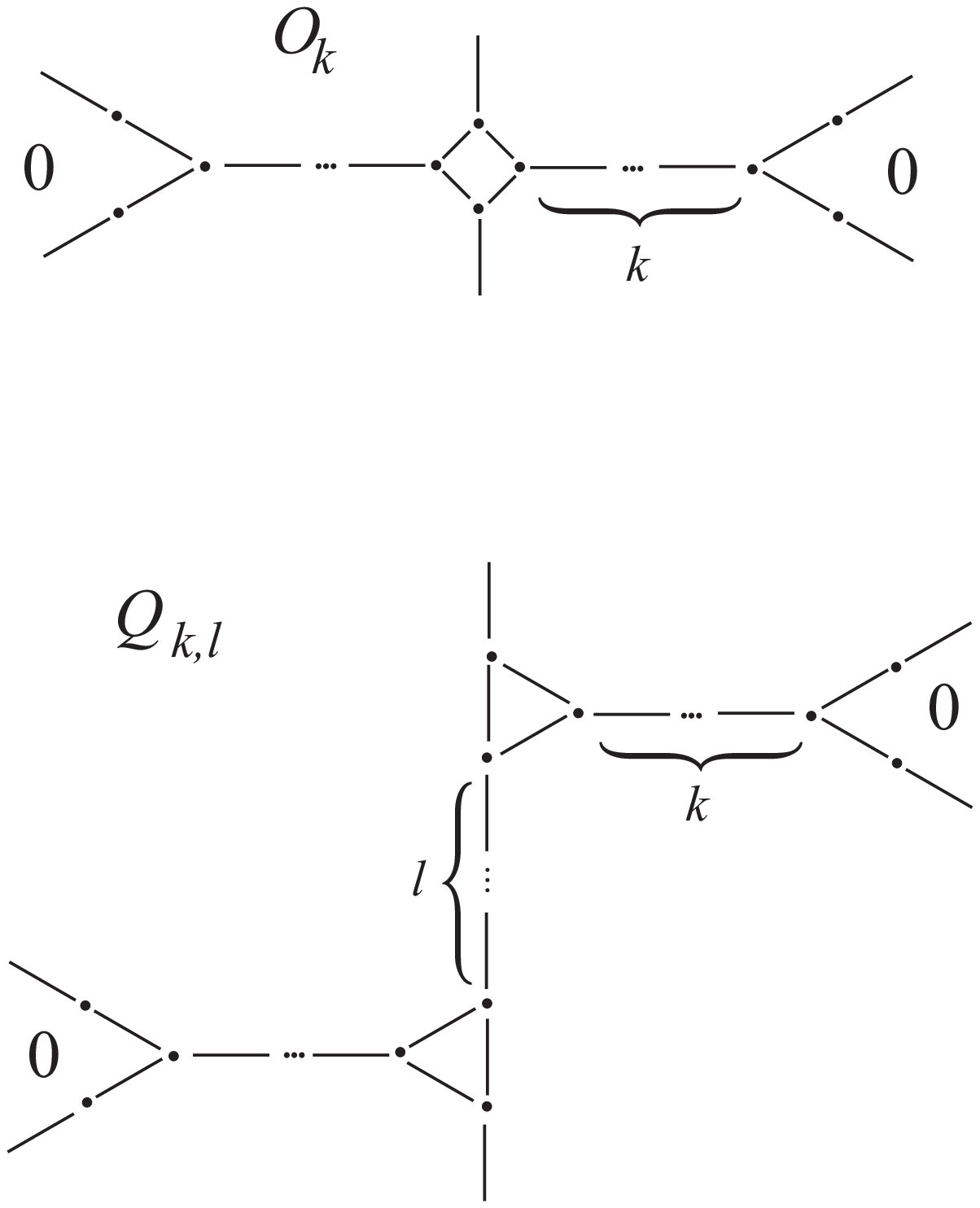}}
\nopagebreak
\vspace{.2in}

\noindent Fig 5. Additional patterns that occur
if the cyclic order of the loops of $\Phi_0$
is opposite to the cyclic order of asymptotic
tracts in $\C$.
\end{center}
\newpage
Propoisition 2 implies that we can continuously deform
the asymptotic values of a function in $G$. We consider
such deformations that the configuration of
5 asymptotic values
describes a closed loop in the space of 5-point
configurations 
symmetric with respect to the origin,  and
the asymptotic values $i$ and $-i$ are interchanged.
Then we compute the action 
of these deformations on our cell
decompositions of the plane, and conclude that
there are exactly two orbits.

\begin{center}
\epsfxsize=3.0in%
\centerline{\epsffile{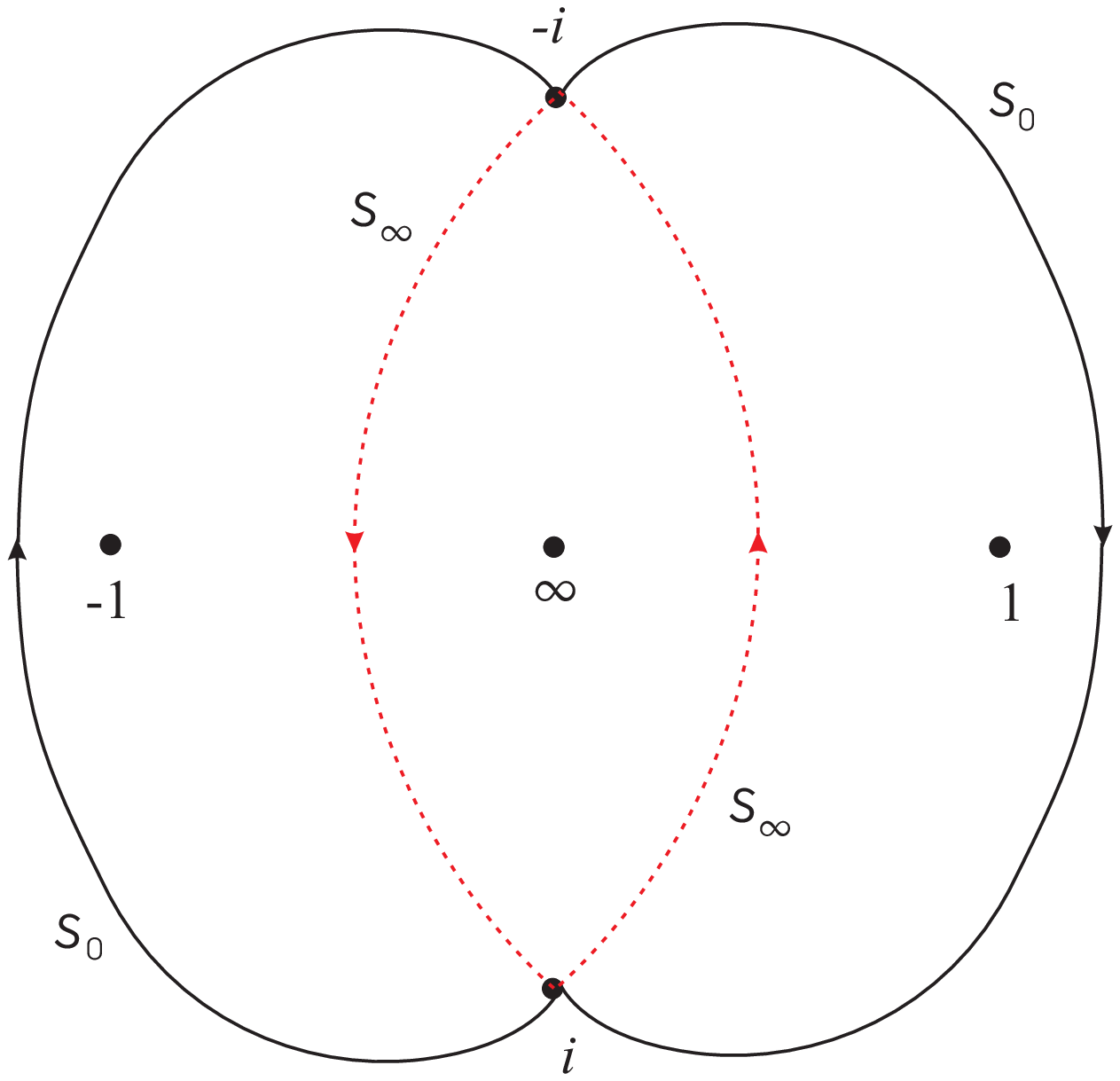}}
\nopagebreak
\vspace{.2in}

\noindent Fig 6. Paths $s_0$ and $s_\infty$ used in the deformation
of $\Psi_0$.
\end{center}
\newpage

\def\g{\gamma}
\begin{center}
\epsfxsize=6.0in%
\centerline{\epsffile{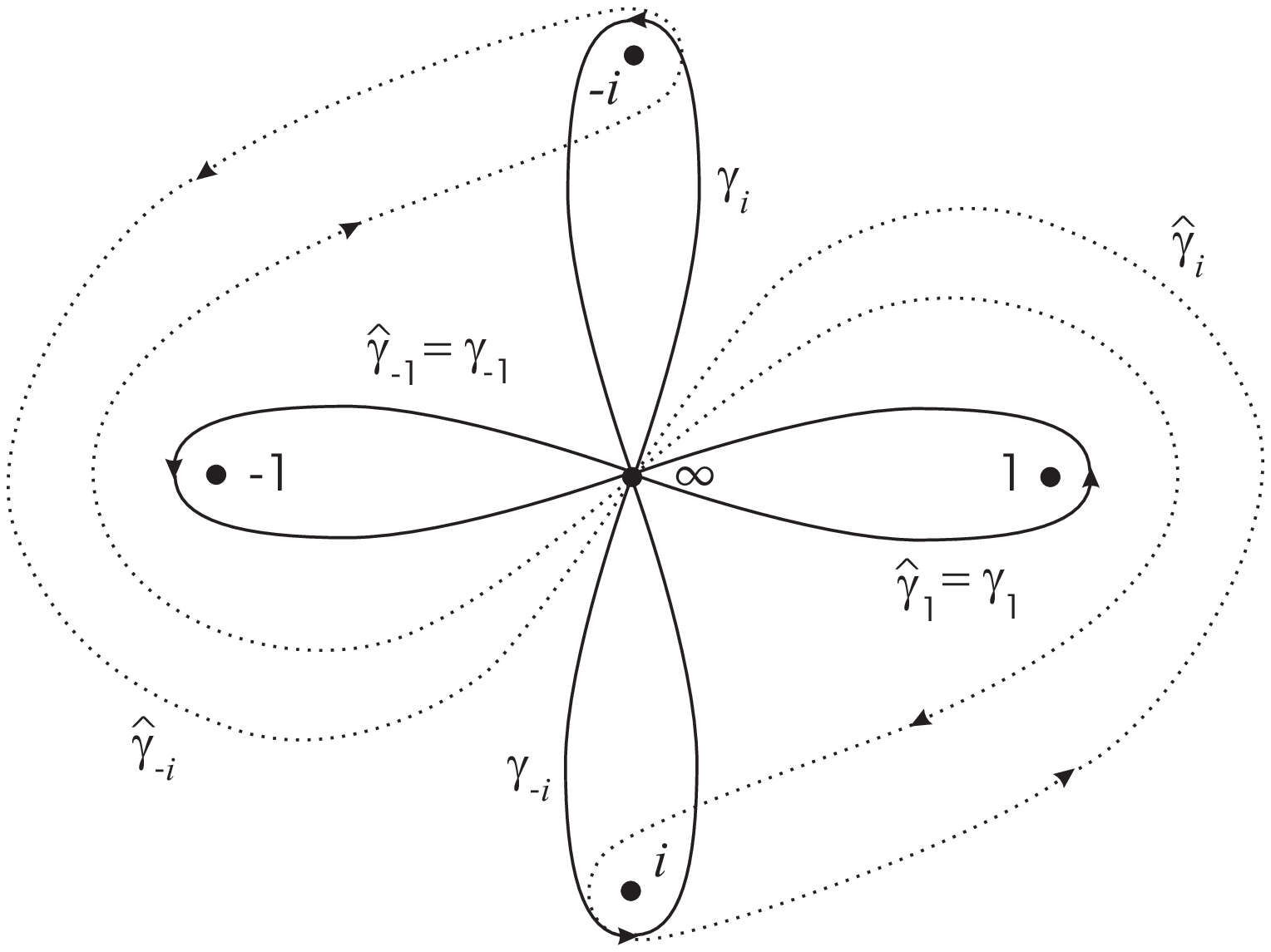}}
\nopagebreak
\vspace{.2in}

\noindent Fig 7. Action of $s_0$ on $\Psi_0$.
The new loops are expressed in terms of the old ones by
the formulas
$$\hat\g_i=(\g_1)^{-1} \g_i \g_1,\quad \hat\g_1=\g_1,$$
$$\hat\g_{-i}=(\g_{-1})^{-1} \g_
{-i} \g_{-1},\quad \hat\g_{-1}=\g_{-1}.$$

\epsfxsize=4.5in%
\centerline{\epsffile{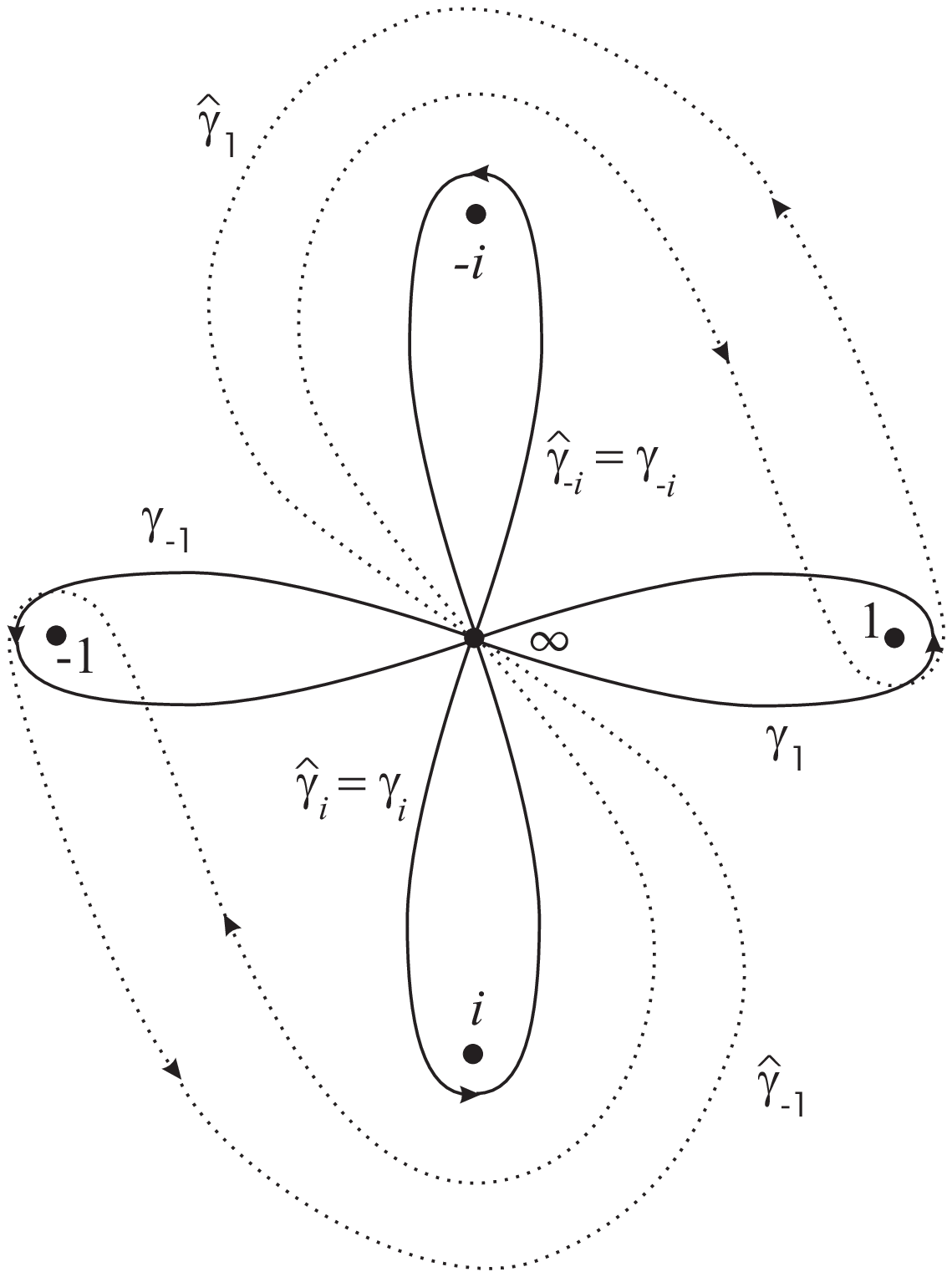}}
\nopagebreak
\vspace{.1in}

\noindent Fig 8. Action of $s_\infty$ on $\Psi_0$.
The new loops are expressed by the formulas
$$\hat\g_i=\g_i,\quad \hat\g_1=(\g_{-i})^{-1} \g_1 \g_{-i},$$
$$\hat\g_{-i}=\g_{-i},\quad
\hat\g_{-1}=(\g_i)^{-1} \g_{-1} \g_i.$$
\newpage

\epsfxsize=5.5in%
\centerline{\epsffile{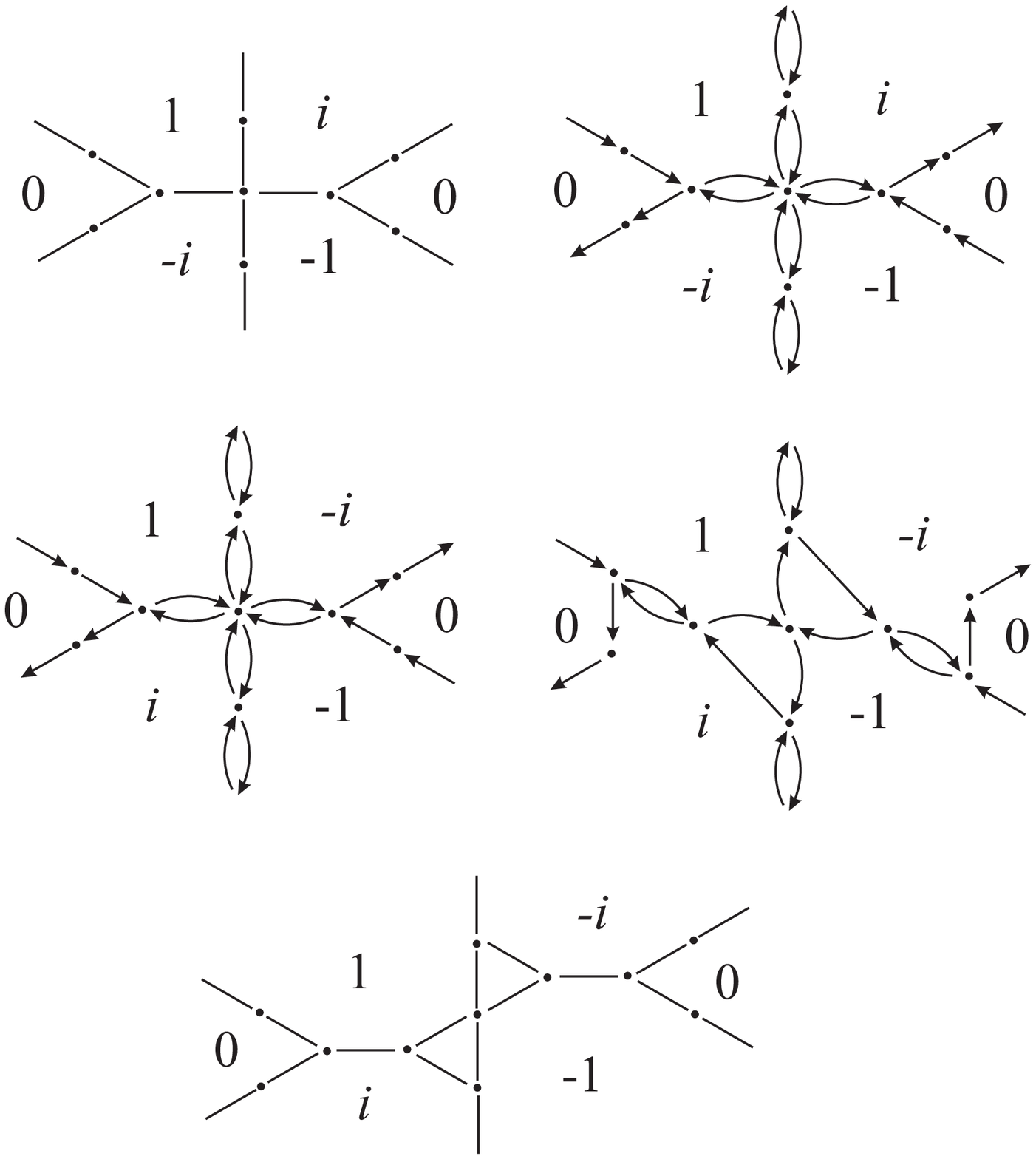}}
\nopagebreak
\vspace{.2in}

\noindent Fig 9. Example. Transformation of $A_1$ to $Q_{1,0}$ by the
action of $s_0$.
\newpage

\epsfxsize=6.3in%
\centerline{\epsffile{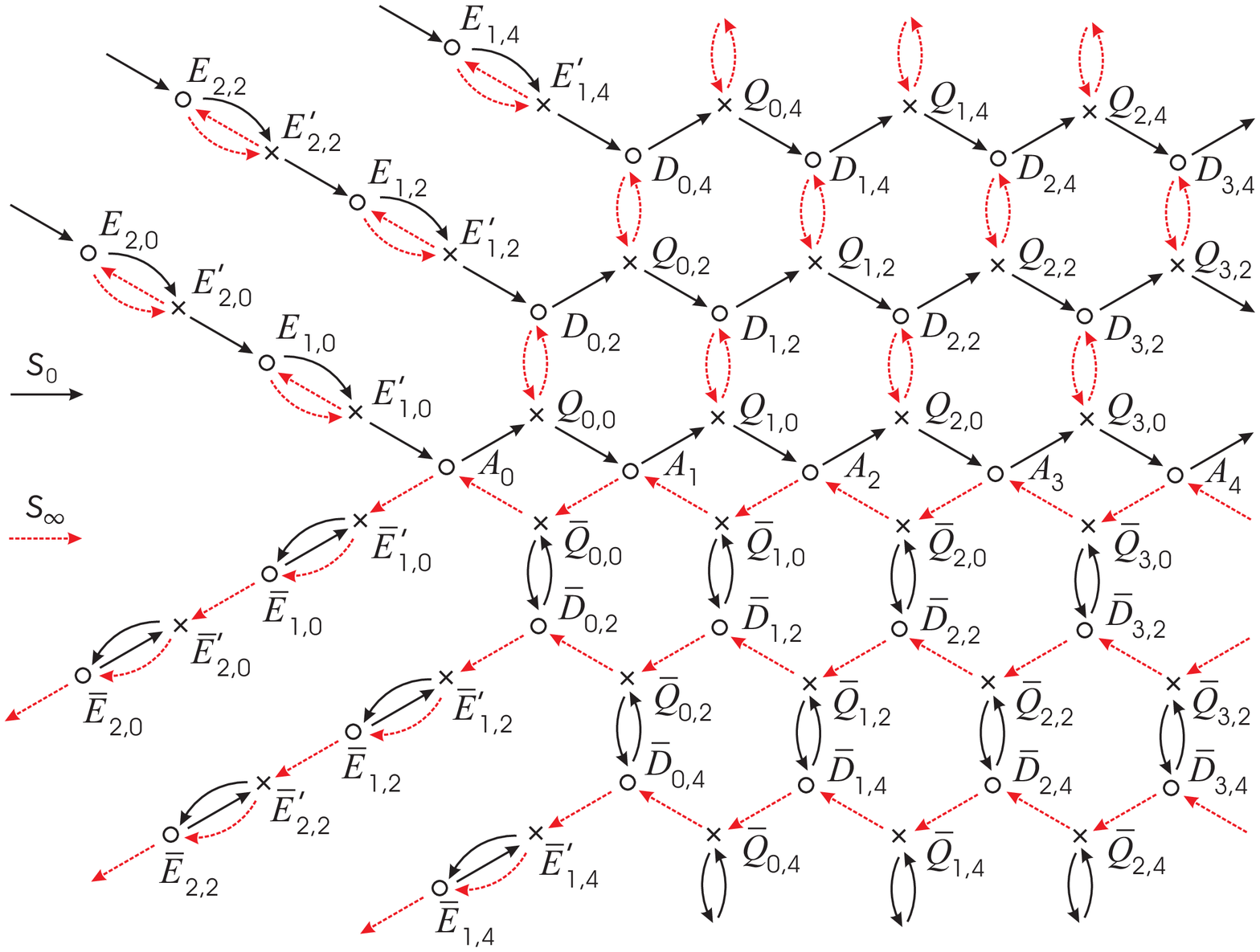}}
\nopagebreak
\vspace{.2in}

\noindent Fig 10. Monodromy action on the even part of $G$.

\epsfxsize=6.3in%
\centerline{\epsffile{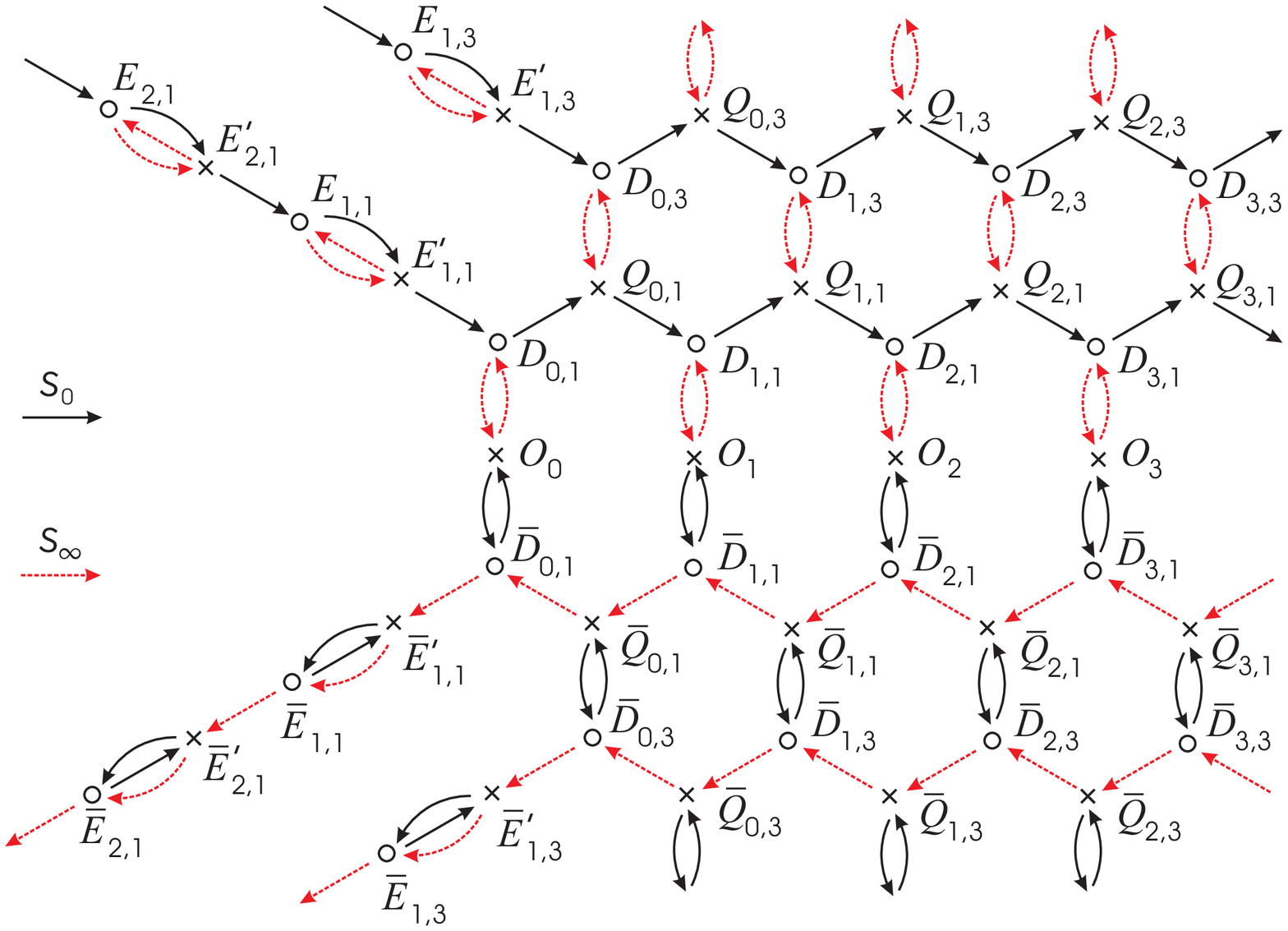}}
\nopagebreak
\vspace{.2in}

\noindent Fig 11. Monodromy action on the odd part of $G$.
\end{center}
\newpage

This proof can be generalized to several other
eigenvalue problems with polynomial potentials
depending on one parameter.

Meromorphic functions in the plane of finite order,
without critical points
are called {\em Nevanlinna functions}.

This class of functions coincides with solutions
of
Schwarz differential equations
$$\frac{f^{\prime\prime\prime}}{f'}-\frac{3}{2}
\left(\frac{f^{\prime\prime}}{f'}\right)^2=P,$$
where $P$ is a polynomial. The order of growth of $f$
equals $({\mathrm{deg}}P)/2+1$, and $f$ has
exactly $({\mathrm{deg}} P)+2$ asymptotic tracts.

The general solution of 
the Schwarz differential equation is a ratio of two
linearly independent solutions of the linear differential
equation 
$$y^{\prime\prime}+\frac{1}{2}Py=0.$$
A Nevanlinna function is determined, up to an affine change
of the independent variable, by its asymptotic values
and by certain combinatorial information---topology of the
cell decomposition of the plane obtained as the
preimage of a cell decomposition of the Riemann
sphere each of whose faces
contains one asymptotic value. Two cell decompositions
of the plane are equivalent if they can be obtained one
from another by a homeomorphism of the
plane preserving orientation.

Our proofs in all cases follow the same pattern.
We parametrize the zero locus
of the spectral determinant by a class of Nevanlinna
functions. To study the irreducible components of
this class, we move their asymptotic values to
a convenient position. Then we describe all Nevanlinna
functions of our class with these asymptotic values
by some cell decompositions of the plane.
All possible cell decompositions arising in a given
problem can be classified by using some related trees.
Then we study the monodromy action on the set of these
trees.
\newpage


\begin{center}
{\bf Quasi-Exactly Solvable sextics} 
(Turbiner-Ushveridze)
\end{center}
\vspace{.1in}
\begin{equation}\label{3}
-y''+(z^6+2\alpha z^4+\{\alpha^2-(4m+2p+3)\}\,z^2)\,y=\lambda y
\end{equation}
\begin{equation}\label{4}
y(\pm\infty)=0\;\mbox{on the real line,}\; m\ge 0,\;p\in\{0,\,1\}.
\end{equation}
There are $m+1$ ``elementary'' eigenfunctions 
$$Q(z)\,\exp(-z^4/4-\alpha z^2/2),$$ 
where $Q$ is a polynomial of degree $2m+p$.

Consider the set $Z_{m,p}$
of pairs $(\alpha,\lambda)$ such that
$\lambda$ is an eigenvalue corresponding to an elementary
eigenfunction of (3), (4).
\vspace{.1in}

\noindent
{\bf Theorem 2} {\em For each $m$ and $p$ the zet $Z_{m,p}$
is irreducible.}
\begin{center}
\epsfxsize=3.0in%
\centerline{\epsffile{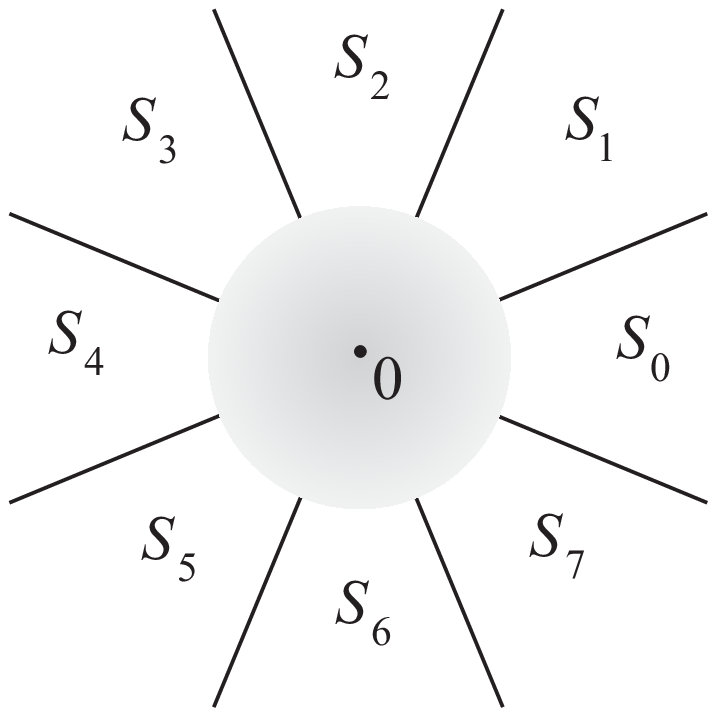}}
\nopagebreak
Fig S1. Stokes sectors for the sextic potential
\end{center}
\vspace{.1in}

Elementary eigenfunctions are distinguished by the
propery that asymptotic values in $S_0,S_2,S_4,S_6$ 
are equal to zero. Thus we have $5$ asymptotic values.
Proposition 2 permits to move them to the points
$(0,i,0,1,0,-i,0,-1)$, and we use the
cell decomposition of the Riemann sphere shown
in Fig. 2. Applying the
reduction procedure illustrated in Fig. 3, 
we obtain centrally symmetric trees with 8 ends,
in which even-numbered faces are labeled by $0$,
and none of such two faces have a common boundary edge.
Next we obtain a 
classification of such trees.
\begin{center}

\epsfxsize=3.5in%
\centerline{\epsffile{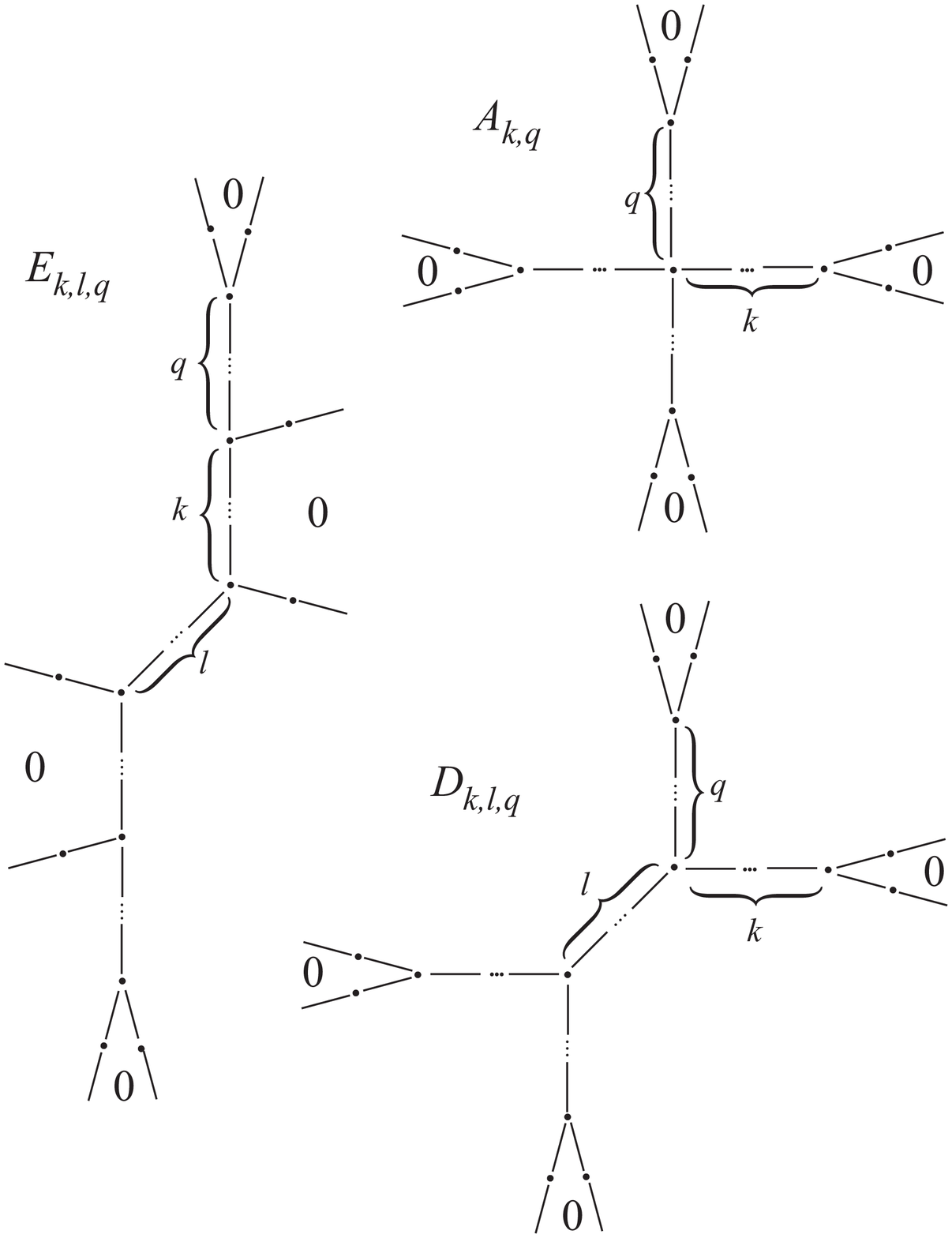}}
\nopagebreak
\vspace{.2in}

\epsfxsize=3.5in%
\centerline{\epsffile{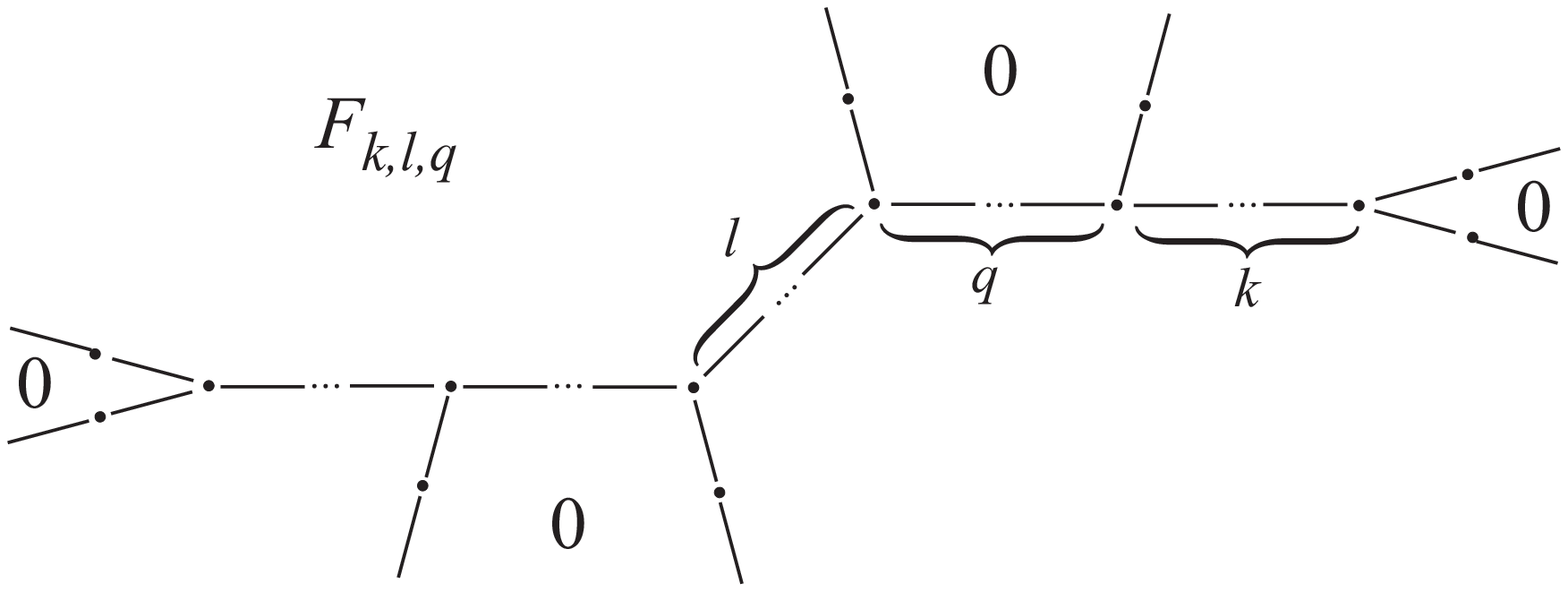}}
\nopagebreak
\vspace{.3in}

\noindent Fig S2. Classification of QES sextic trees
\newpage
\epsfxsize=5.0in%
\centerline{\epsffile{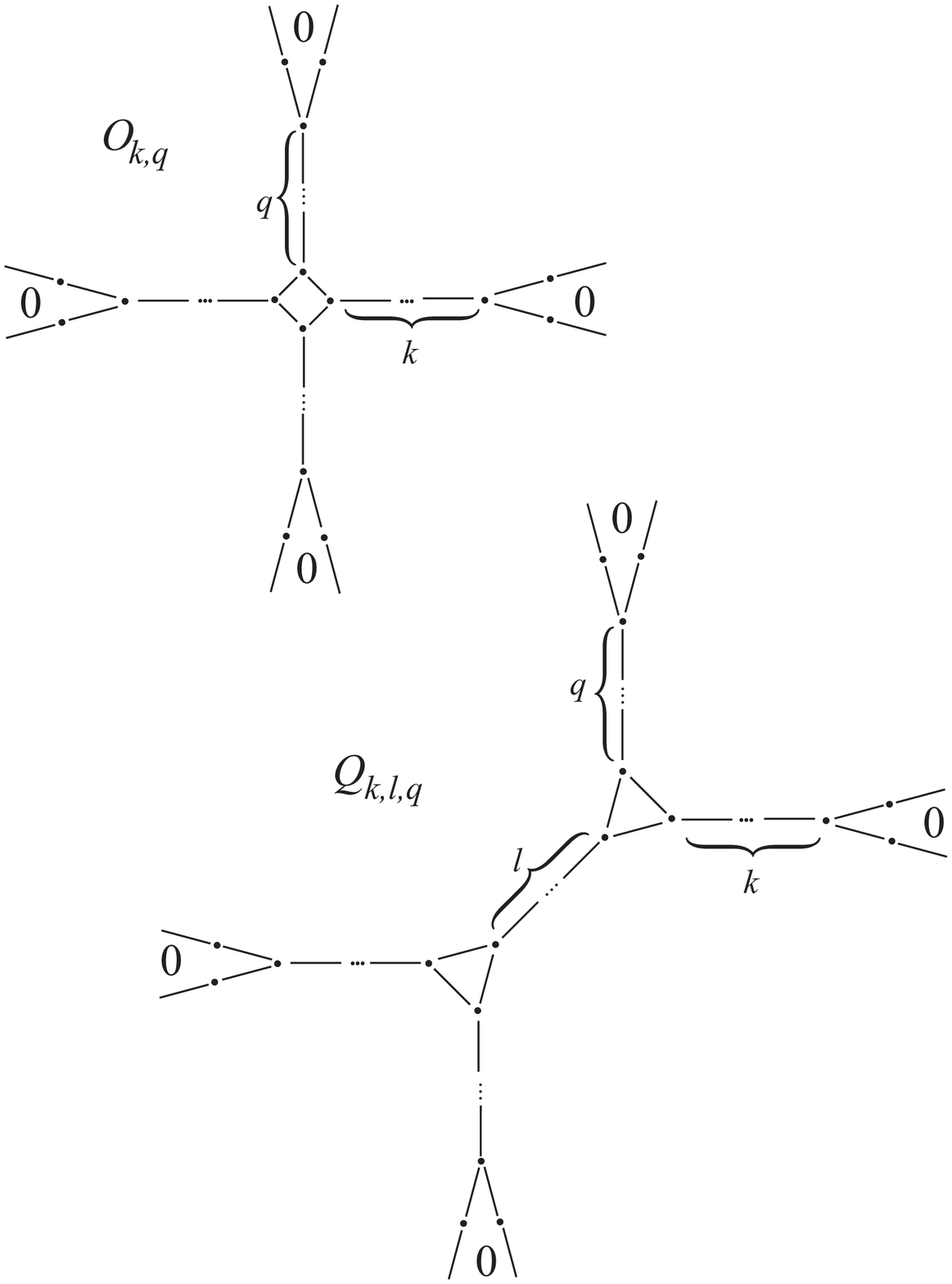}}
\nopagebreak
\vspace{.2in}

\noindent Fig S3. Non-tree QES sextic graphs

\newpage
\epsfxsize=4.8in%
\centerline{\epsffile{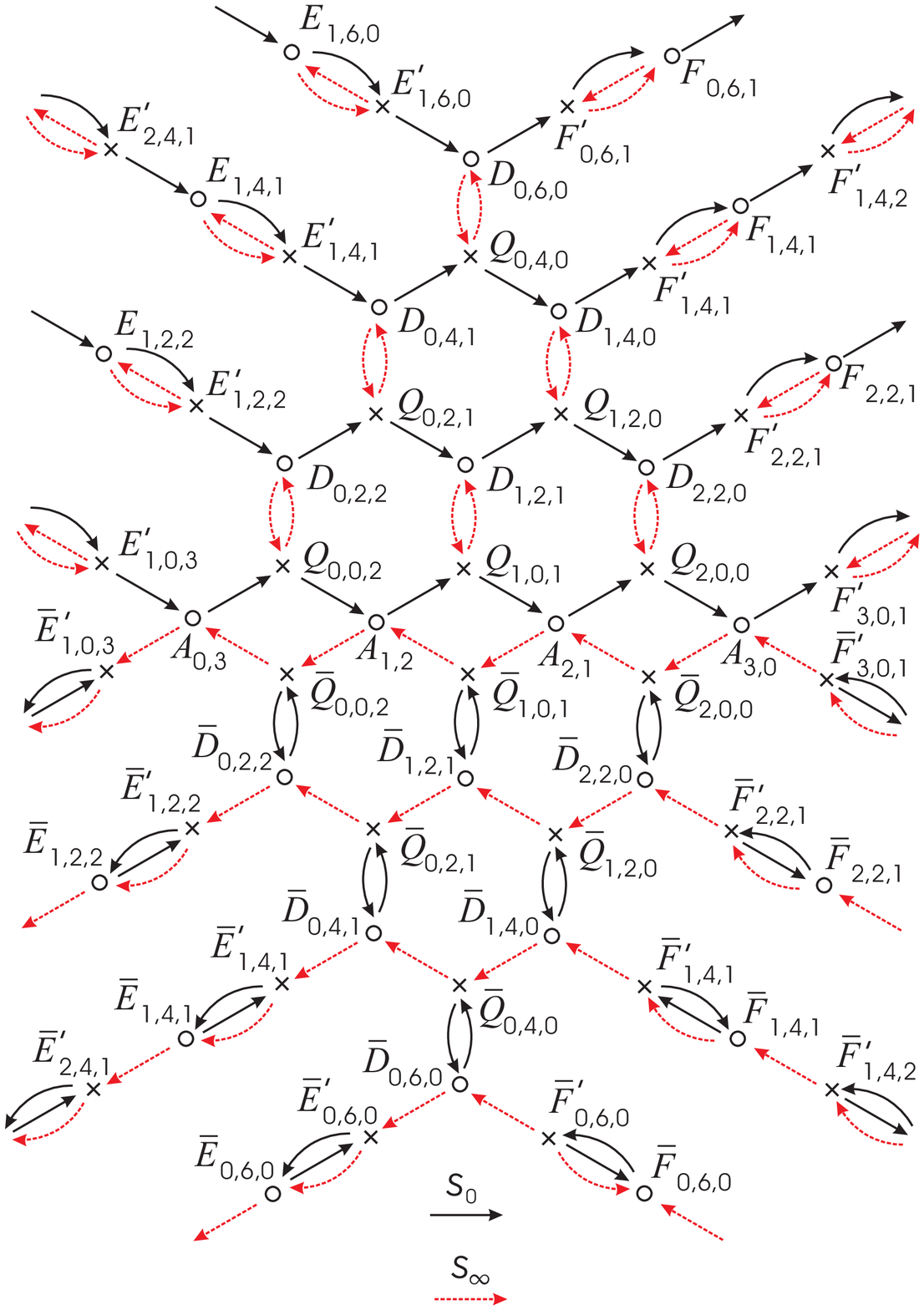}}
\nopagebreak
\vspace{.1in}
\noindent Fig S4. Monodromy action for an even QES sextic, $m=3,\;p=0$
\newpage
\epsfxsize=5.5in%
\centerline{\epsffile{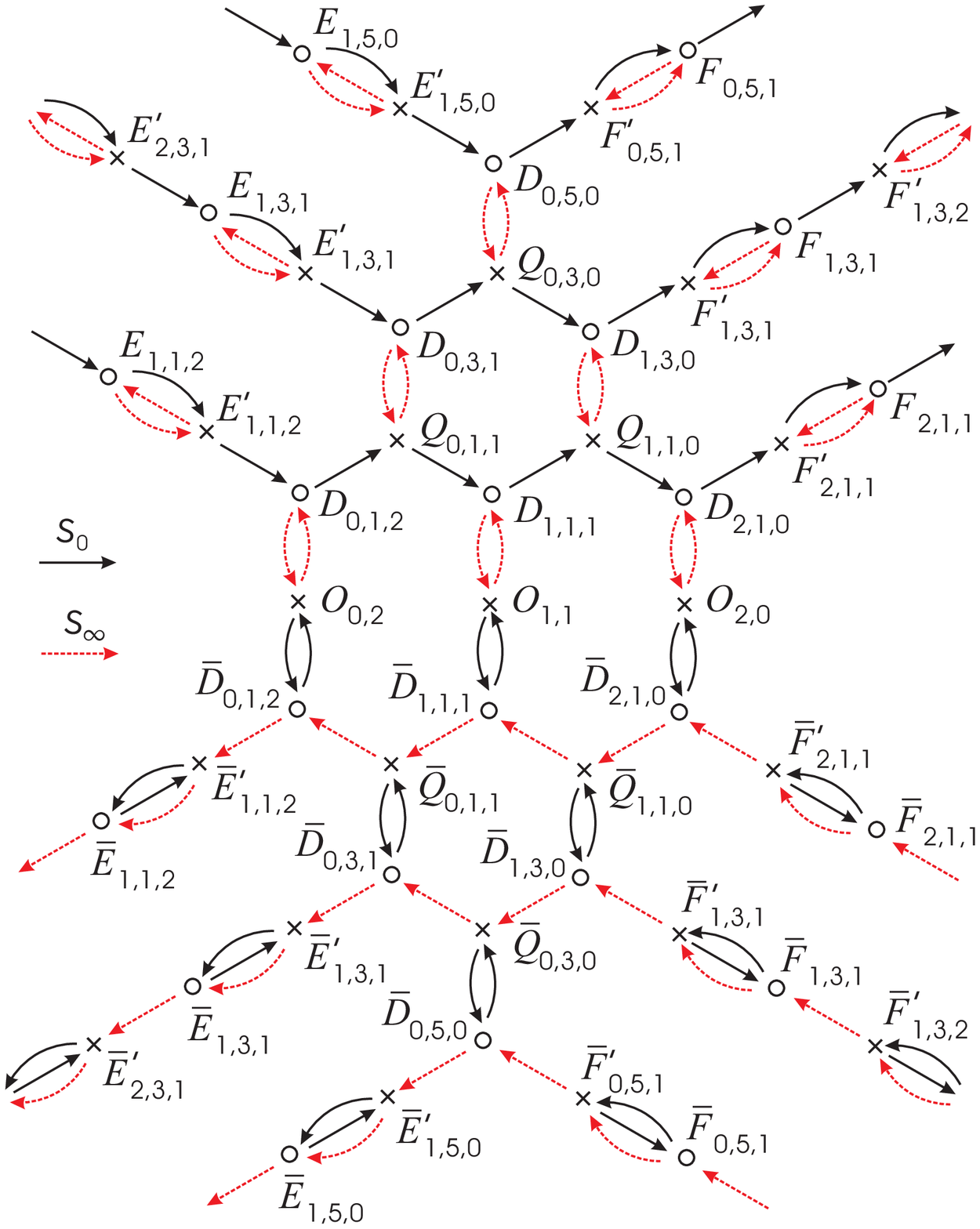}}
\nopagebreak
\vspace{.2in}

\noindent Fig S5. Monodromy action for an odd QES sextic, $m=2,\;p=1$
\newpage
\epsfxsize=4.5in%
\centerline{\epsffile{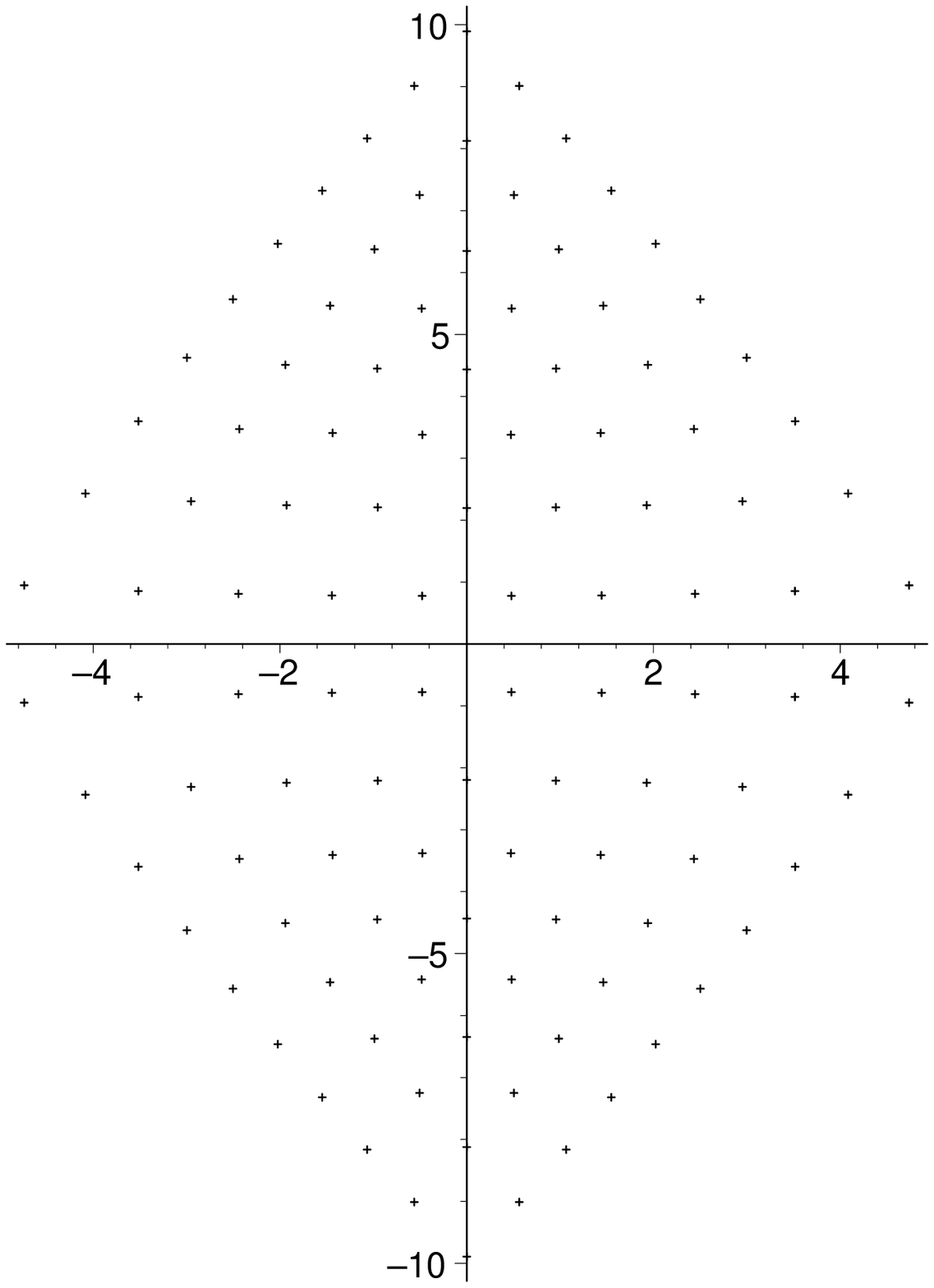}}
\nopagebreak
\vspace{.2in}

\noindent Fig S6. Branching points for a QES sextic with $m=10,\;p=0$
\end{center}
\newpage
\begin{center} \bf{Rescaling}\end{center}
\vspace{.1in}
Let $n=4m+2p+3$. Then the quasi-exactly solvable equation (3) 
is related to
\begin{equation}\label{5}
-y''(x)+[a^2x^6+2abx^4+(b^2-an)x^2]y(x)=\mu y(x)
\end{equation}
by the scaling $z=a^{1/4}x,\; b=a^{1/2}\alpha,\;\lambda=a^{1/2}\mu$.
To approximate the quartic potential $2x^4+\beta x^2$ 
by the rescaled quasi-exactly solvable sextic potentials in (5) 
as $m\to\infty$, let
$b=n^{1/3}(1+sn^{-2/3}),\; a=n^{-1/3}(1+tn^{-2/3})$.
Then $\alpha=b/a^{1/2}=n^{1/2}(1+(s-t/2)n^{-2/3}+O(n^{-4/3})$.
Substituting expression for $a$ and $b$ into (5), we get
the potential
$$
n^{-2/3}(1+O(n^{-2/3}))x^6+2(1+(s+t)n^{-2/3}+stn^{-4/3})x^4
+((2s-t)+s^2 n^{-2/3})x^2.
$$
Hence $\beta=2s-t=2(n^{-1/2}\alpha-1)n^{2/3}+O(n^{-2/3})$.
\newpage
\epsfxsize=4.5in%
\centerline{\epsffile{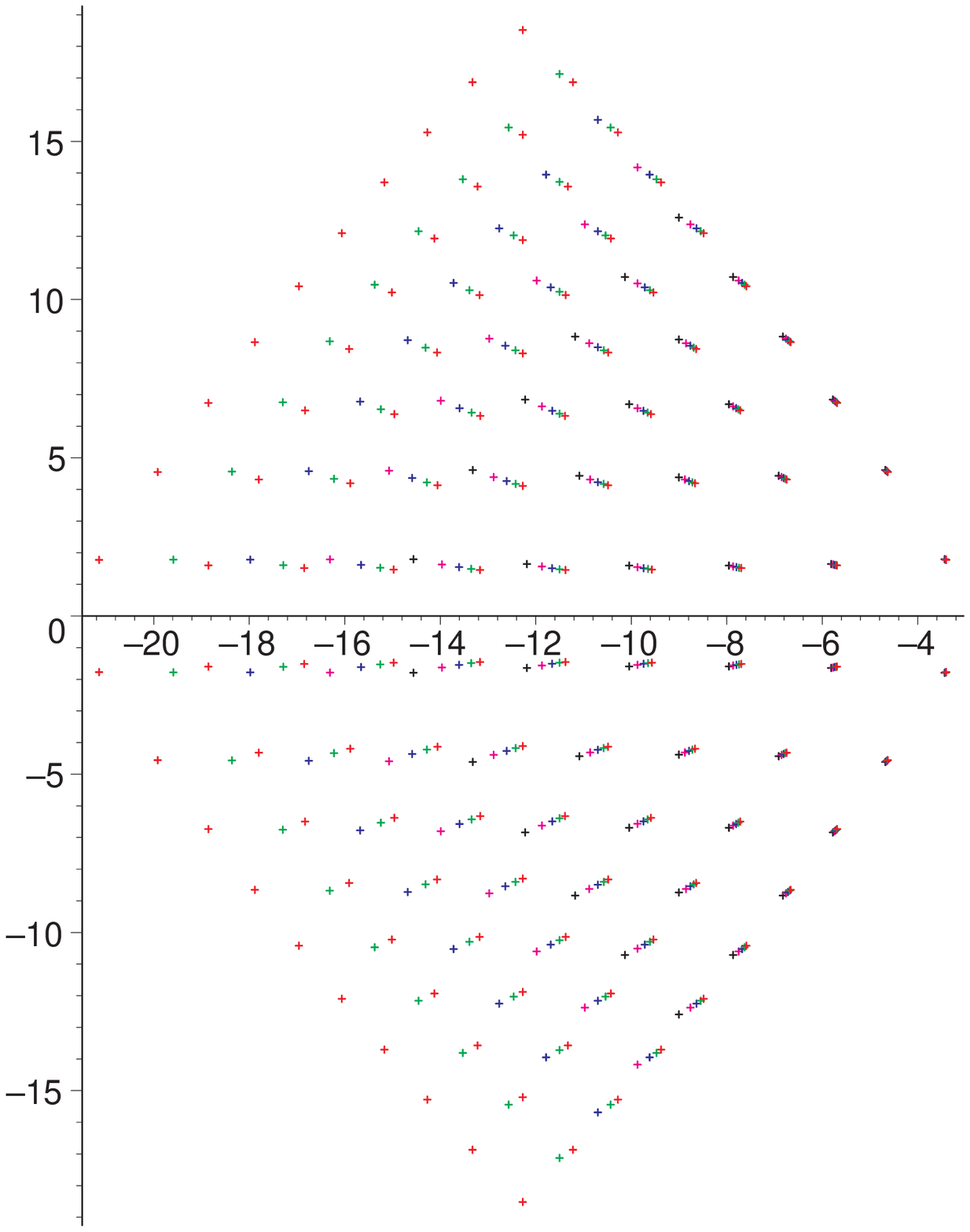}}
\nopagebreak
\vspace{.2in}
 
\begin{center}

\noindent Fig S7. Branching points for rescaled QES sextics 
with $m=6\mbox{-}10,\;p=0$
\newpage
\epsfxsize=4.5in%
\centerline{\epsffile{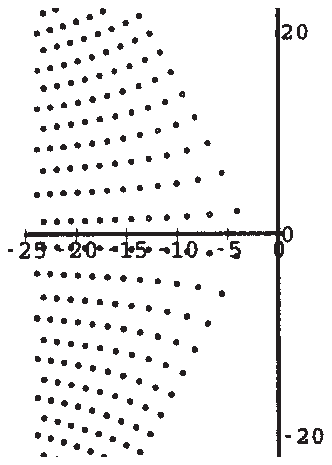}}
\nopagebreak
\vspace{.2in}

\noindent Fig S8. Branching points for the quartic oscillator
(Delabaere--Pham). 
\end{center}
%
\newpage
\begin{center}
{\bf PT-symmetric cubic} (Delabaere-Trinh)
\end{center}
\vspace{.1in}
Polynomial potential $P(z)$ is $PT$-symmetric if $P(-\bar z)=\overline{P(z)}$.

Eigenvalue problem
\begin{equation}\label{6}
-y''+(iz^3+i\alpha z)\,y=\lambda y,
\end{equation}
\begin{equation}\label{7}
y(\pm\infty)=0\;\mbox{on the real line},
\end{equation}
is $PT$-symmetric for real $\alpha$.

Let $Z$ be the set of all pairs $(\alpha,\lambda)$ such that
$\lambda$ is an eigenvalue of (6), (7).
\vspace{.1in}

\noindent
{\bf Theorem 3} {\em The set $Z$ is irreducible}.
\begin{center}
\epsfxsize=3.0in%
\centerline{\epsffile{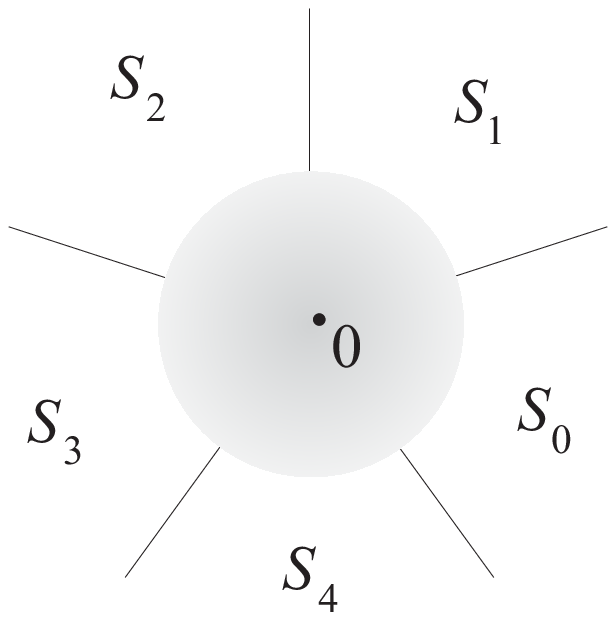}}
\nopagebreak
\vspace{.2in}

\noindent Fig C0. Stokes sectors for the cubic potential
\vspace{.4in}
\newpage
\centerline{\bf Line complexes}
\vspace{.2in}

\end{center}

Let $f$ be a Nevanlinna function with $q$ asymptotic values.
Consider a cell decomposition of the sphere with two
vertices which we denote by $\times$ and $\circ$, with
$q$ edges each connecting these two vertices,
and such that each of the faces contains
exactly one asymptotic value. 
The $f$-preimage of such cell decomposition is called a line complex
(see \cite{GO})
All possible line complexes of Nevanlinna functions
can be simply characterized:
they are bipartite graphs embedded in the plane
such that the degree of each vertex is $q$ and all faces
have either two or infinitely many boundary edges.
Replacing multiple edges of a cell complex by single edges
we obtain a tree. The line complex can be recovered from
this tree. 

These properties make line complexes very convenient.
The reason why we used other cell decompositions of
the sphere for quartics and sextics is that there are no
line complexes for these cases having all symmetries
present in the problems.

In the case of PT-symmetric cubic we can move the asymptotic
values to the following positions: $(0,1,-1,0,\infty)$
(listed counterclockwise starting from $S_0$.
The line complex is the preimage
of the following cell
decomposition of the Riemann sphere: 
\vspace{.1in}

\epsfxsize=4.0in%
\centerline{\epsffile{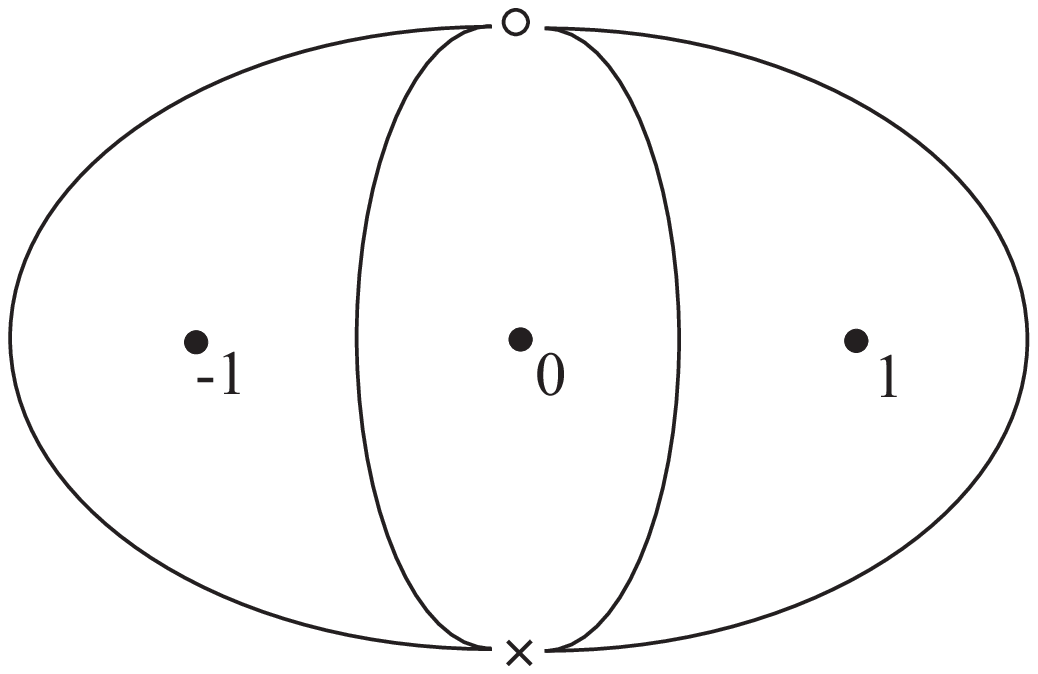}}
\nopagebreak
\vspace{.1in}

\begin{center}
\noindent Fig C1. Cell decomposition of the sphere for the cubic

\epsfxsize=5.0in%
\centerline{\epsffile{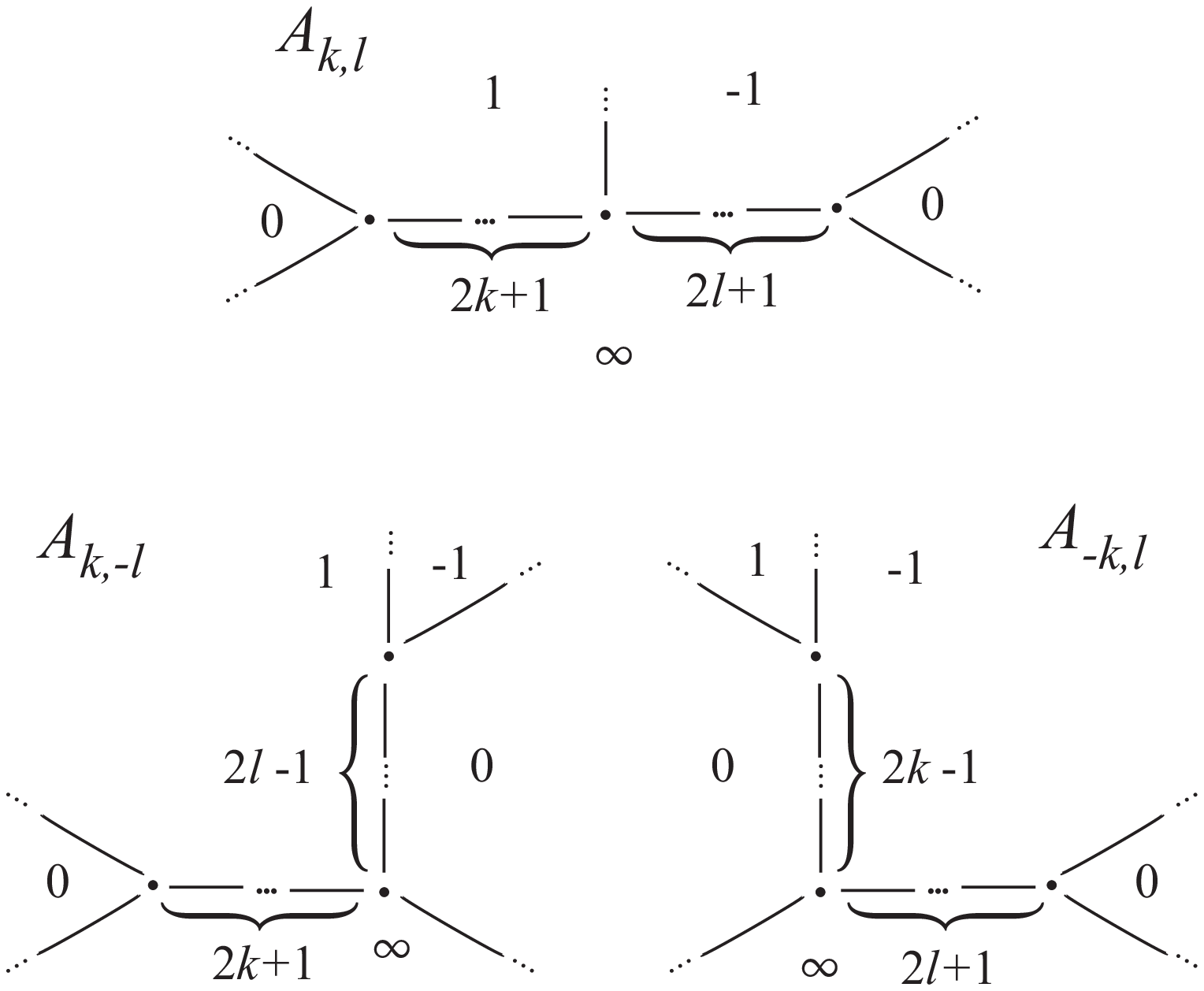}}
\nopagebreak
\vspace{.3in}

\noindent Fig C2. Trees for the cubic. Type $A$
\end{center}
\newpage

\epsfxsize=5.0in%
\centerline{\epsffile{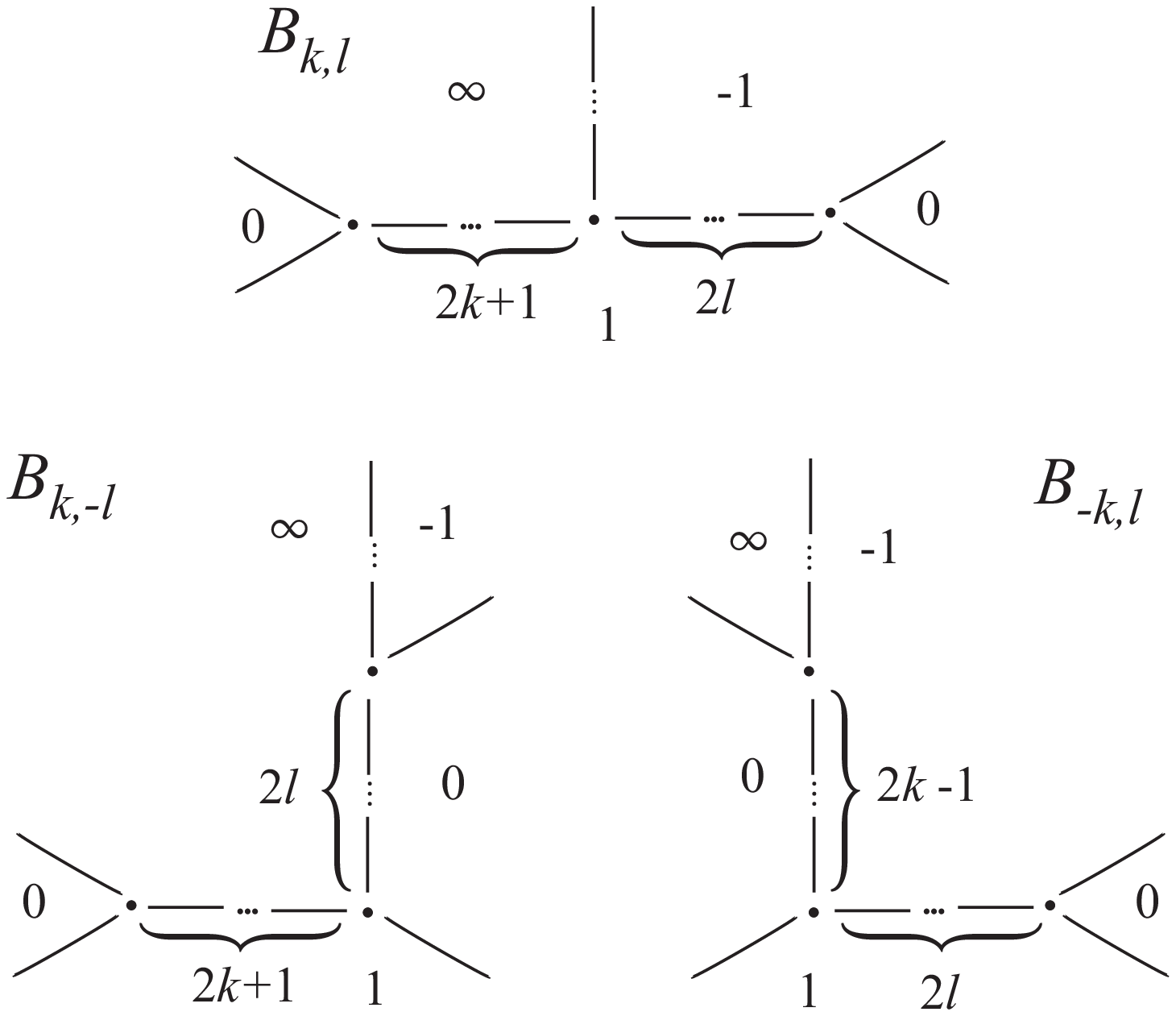}}
\nopagebreak
\vspace{.2in}

\begin{center}
\noindent Fig C3. Trees for the cubic. Type $B$
\newpage

\epsfxsize=5.0in%
\centerline{\epsffile{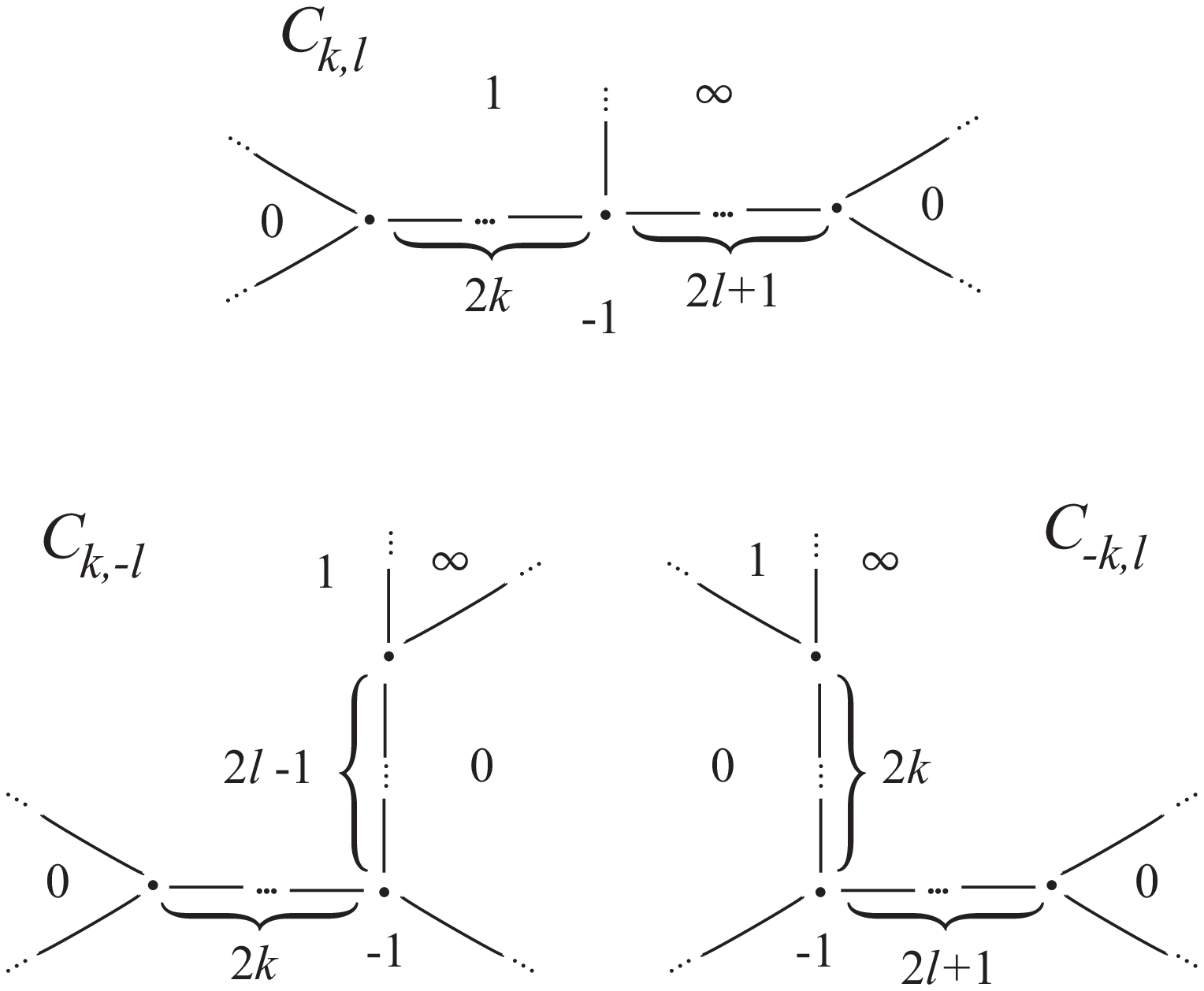}}
\nopagebreak
\vspace{.2in}

\noindent Fig C4. Trees for the cubic. Type $C$
\newpage

\epsfxsize=3.0in%
\centerline{\epsffile{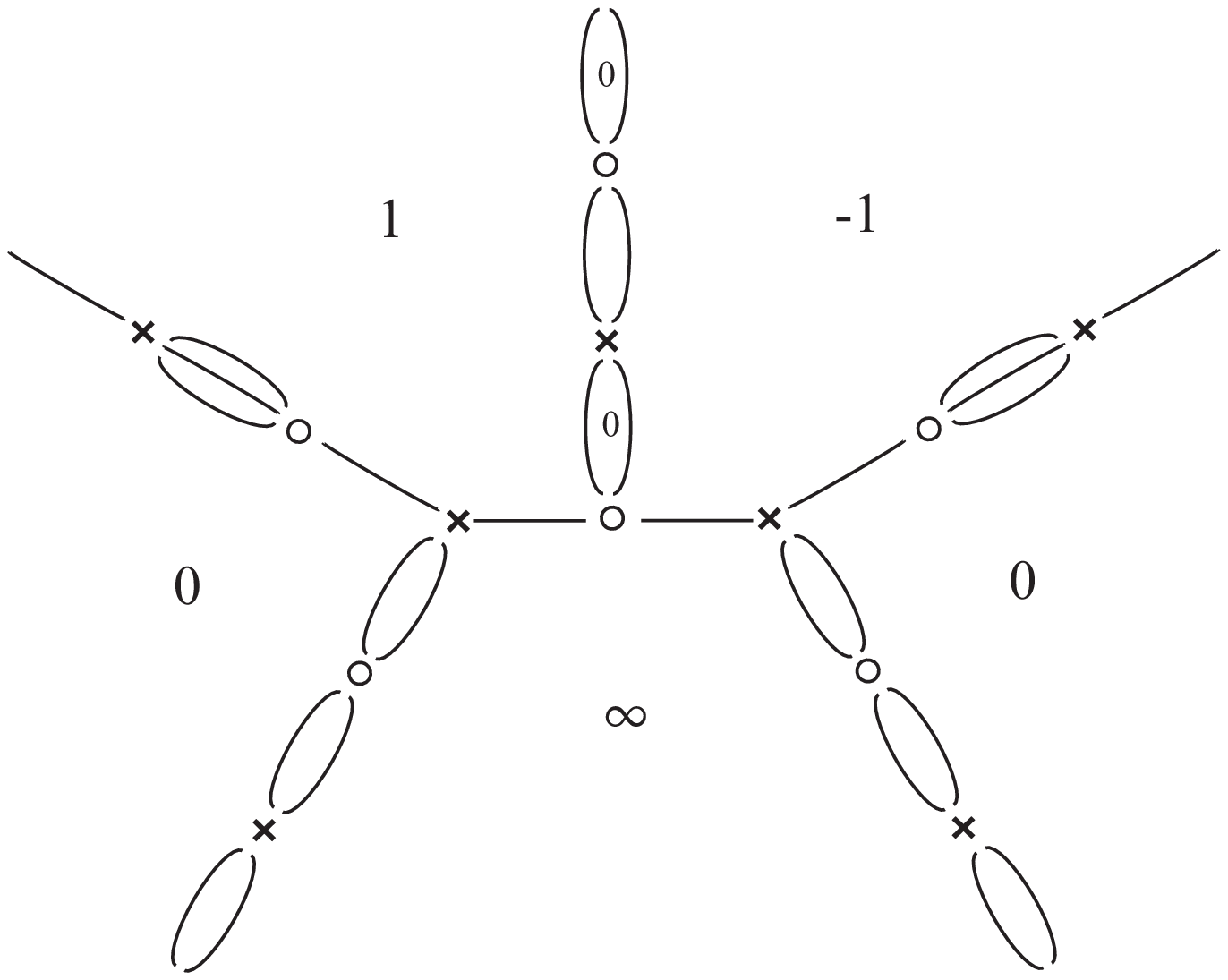}}
\nopagebreak
\vspace{.2in}

\noindent Fig C5. Line complex of type $A_{0,0}$
\vspace{.4in}

\epsfxsize=3.0in%
\centerline{\epsffile{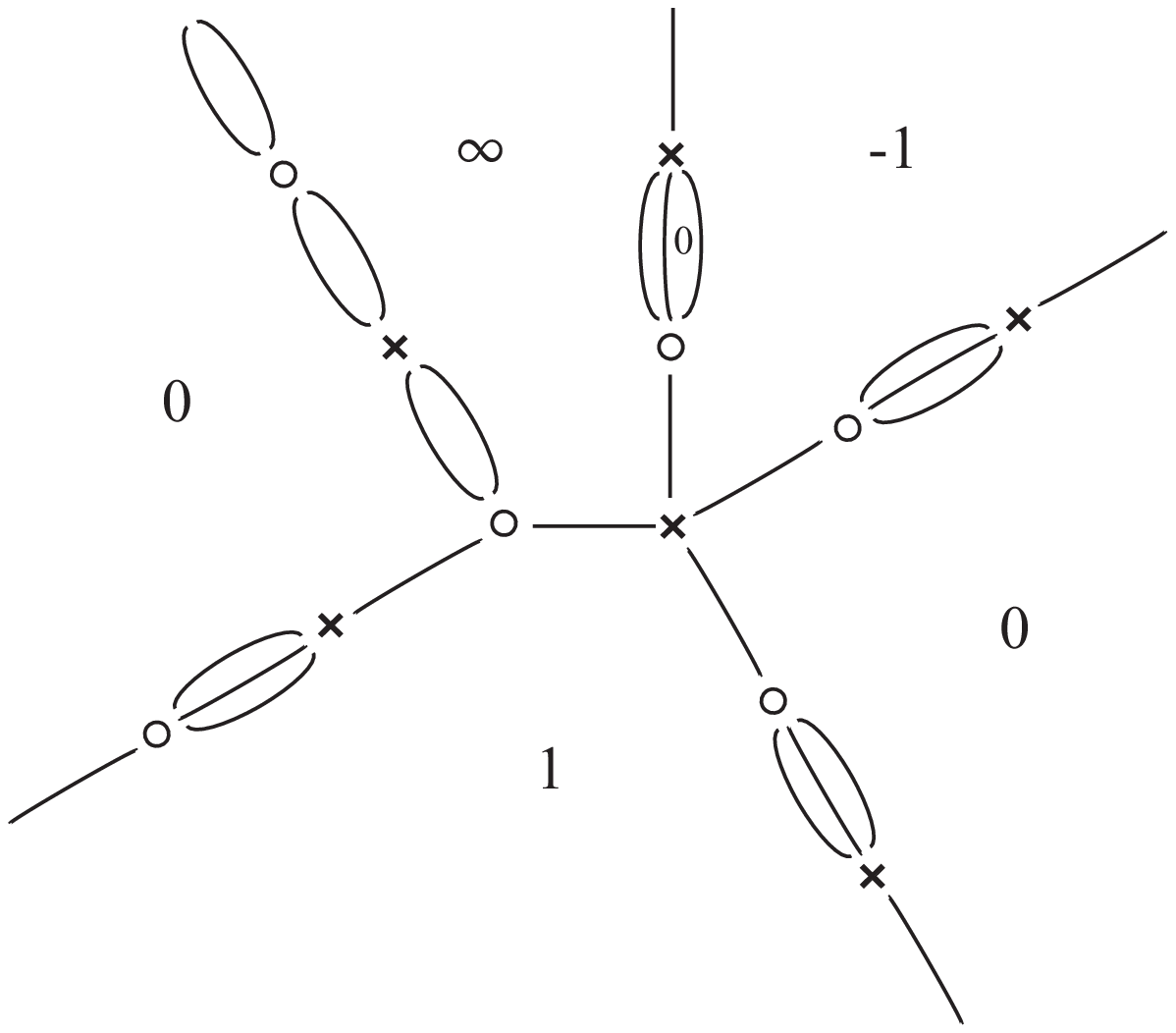}}
\nopagebreak
\vspace{.2in}

\noindent Fig C6. Line complex of type $B_{0,0}$

\epsfxsize=3.0in%
\centerline{\epsffile{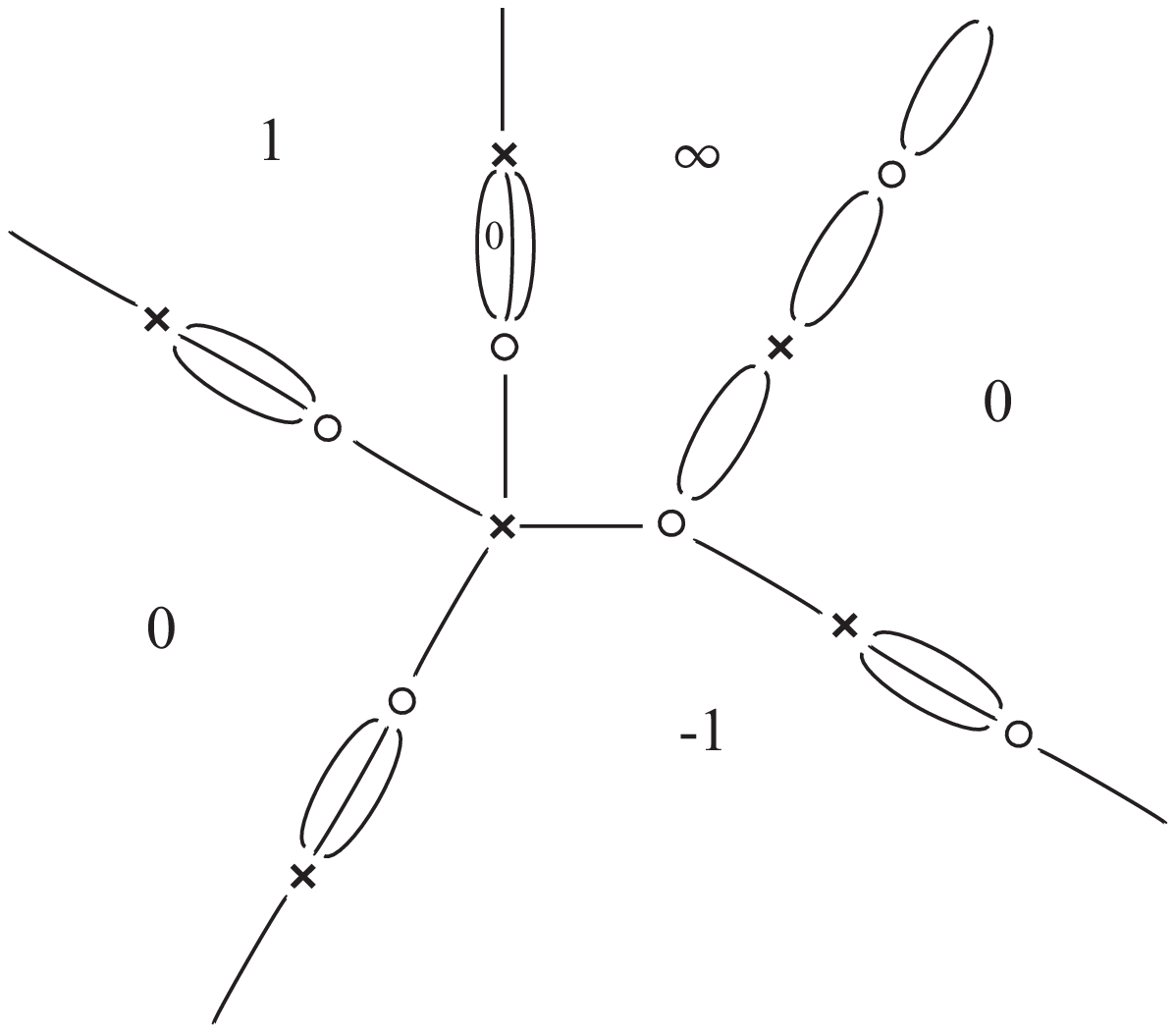}}
\nopagebreak
\vspace{.2in}

\noindent Fig C7. Line complex of type $C_{0,0}$
\vspace{.4in}

\epsfxsize=3.0in%
\centerline{\epsffile{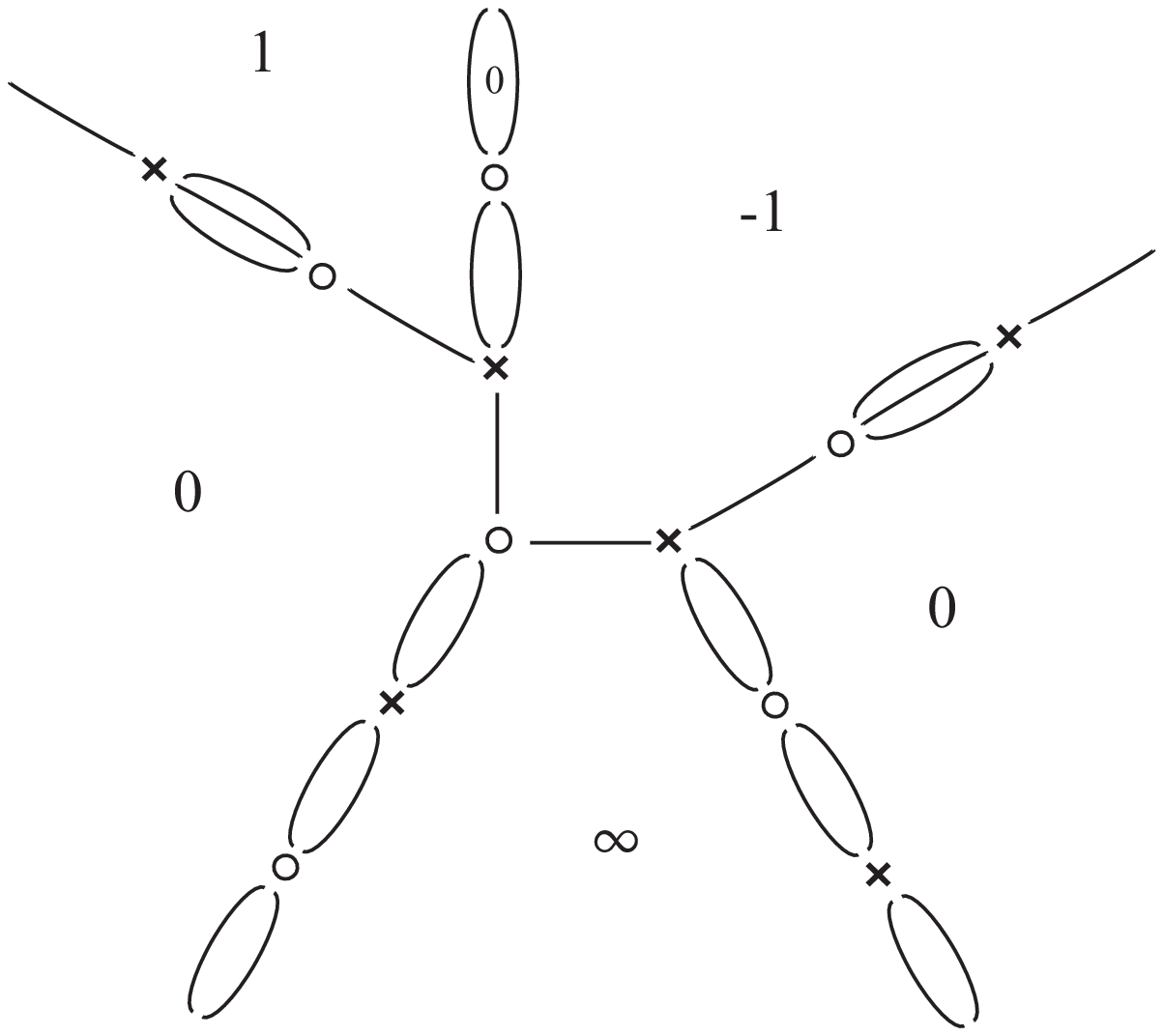}}
\nopagebreak
\vspace{.2in}

\noindent Fig C8. Line complex of type $A_{-1,0}$

\epsfxsize=3.0in%
\centerline{\epsffile{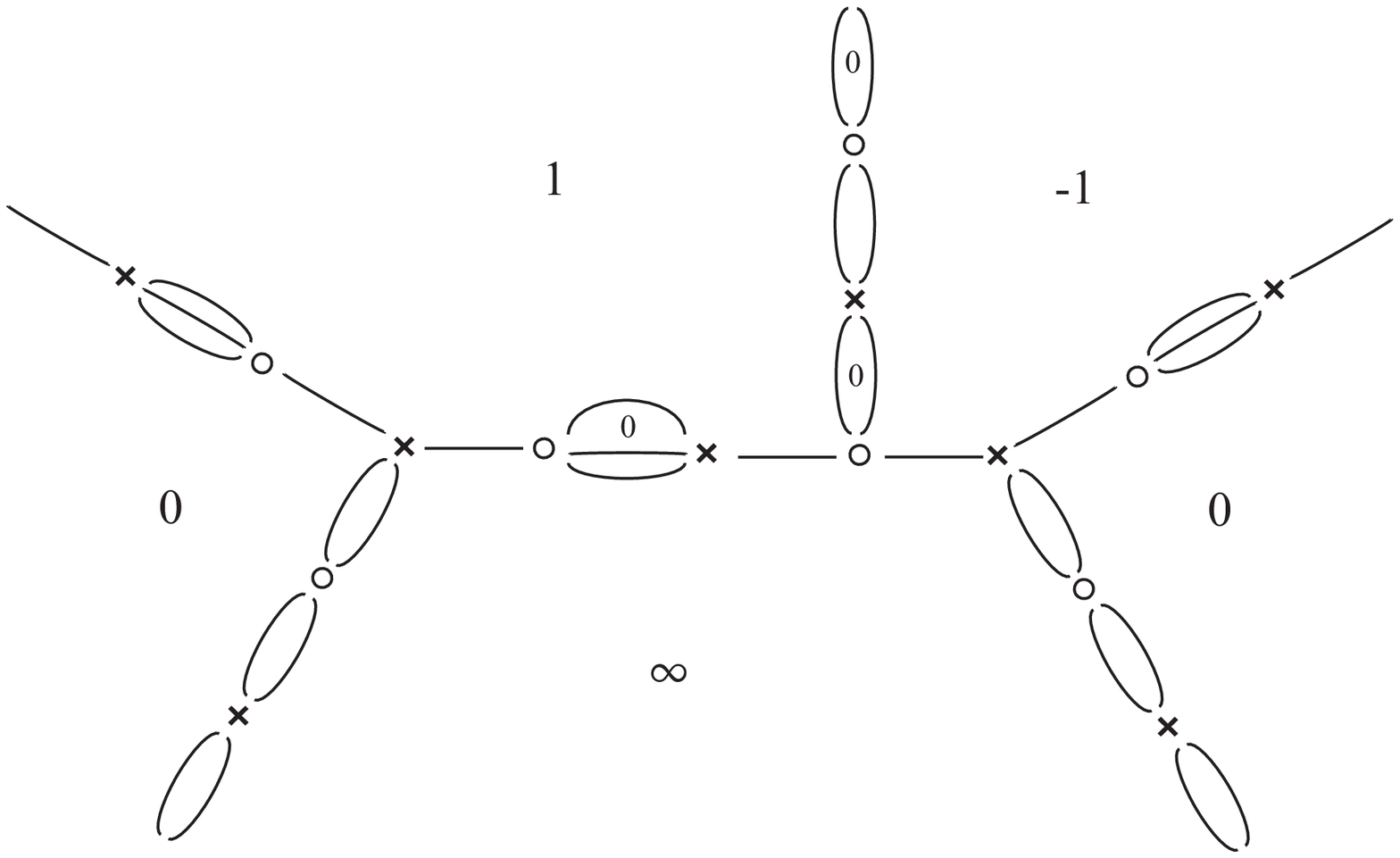}}
\nopagebreak
\vspace{.2in}

\noindent Fig C9. Line complex of type $A_{1,0}$
\vspace{.2in}

\epsfxsize=4.0in%
\centerline{\epsffile{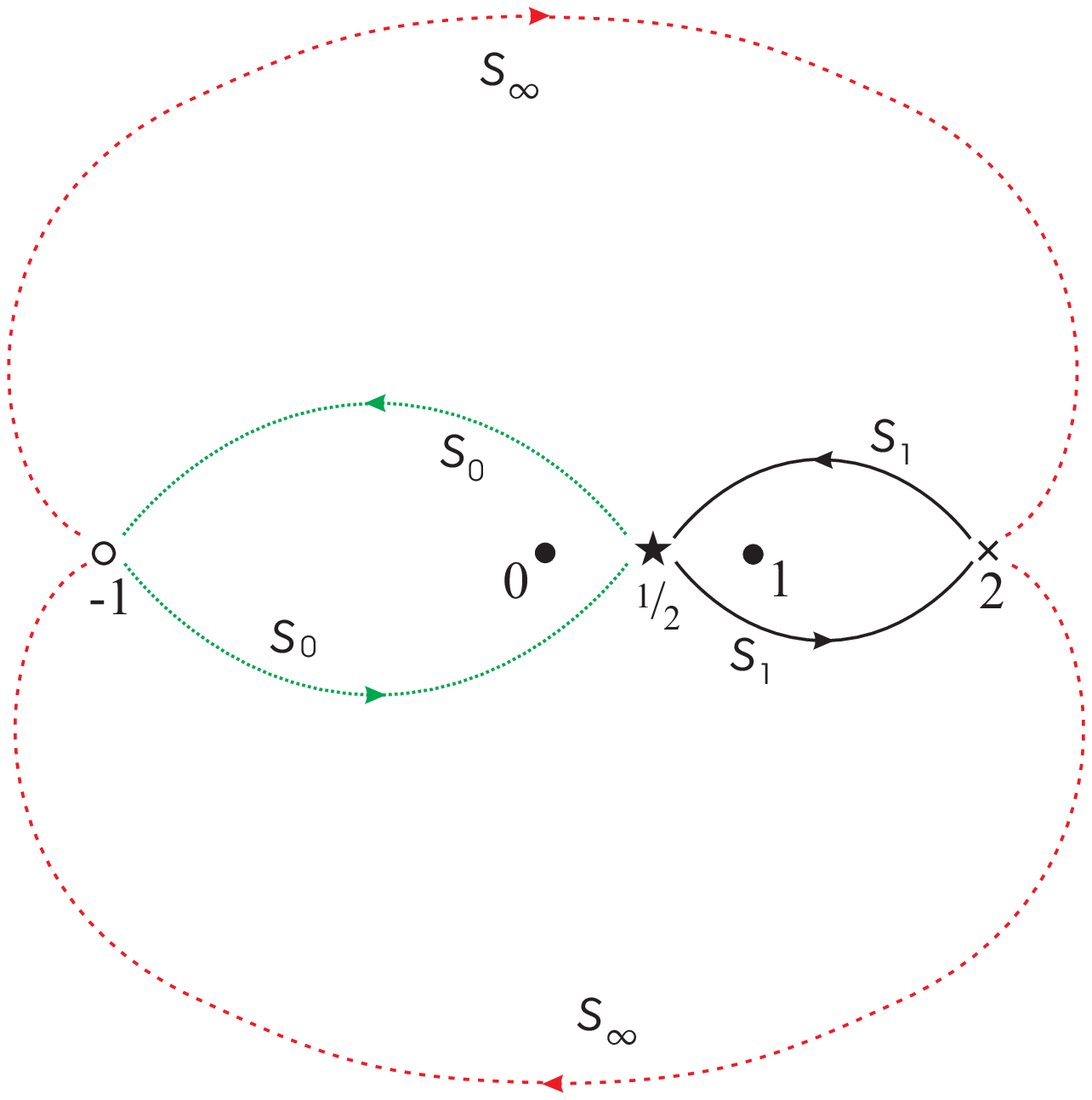}}
\nopagebreak
\vspace{.2in}
\noindent Fig C10. Paths $s_0,\; s_1$ and $s_\infty$
used in the deformation of the Nevanlinna function
corresponding to the cubic.
\end{center}

\epsfxsize=6.0in%
\centerline{\epsffile{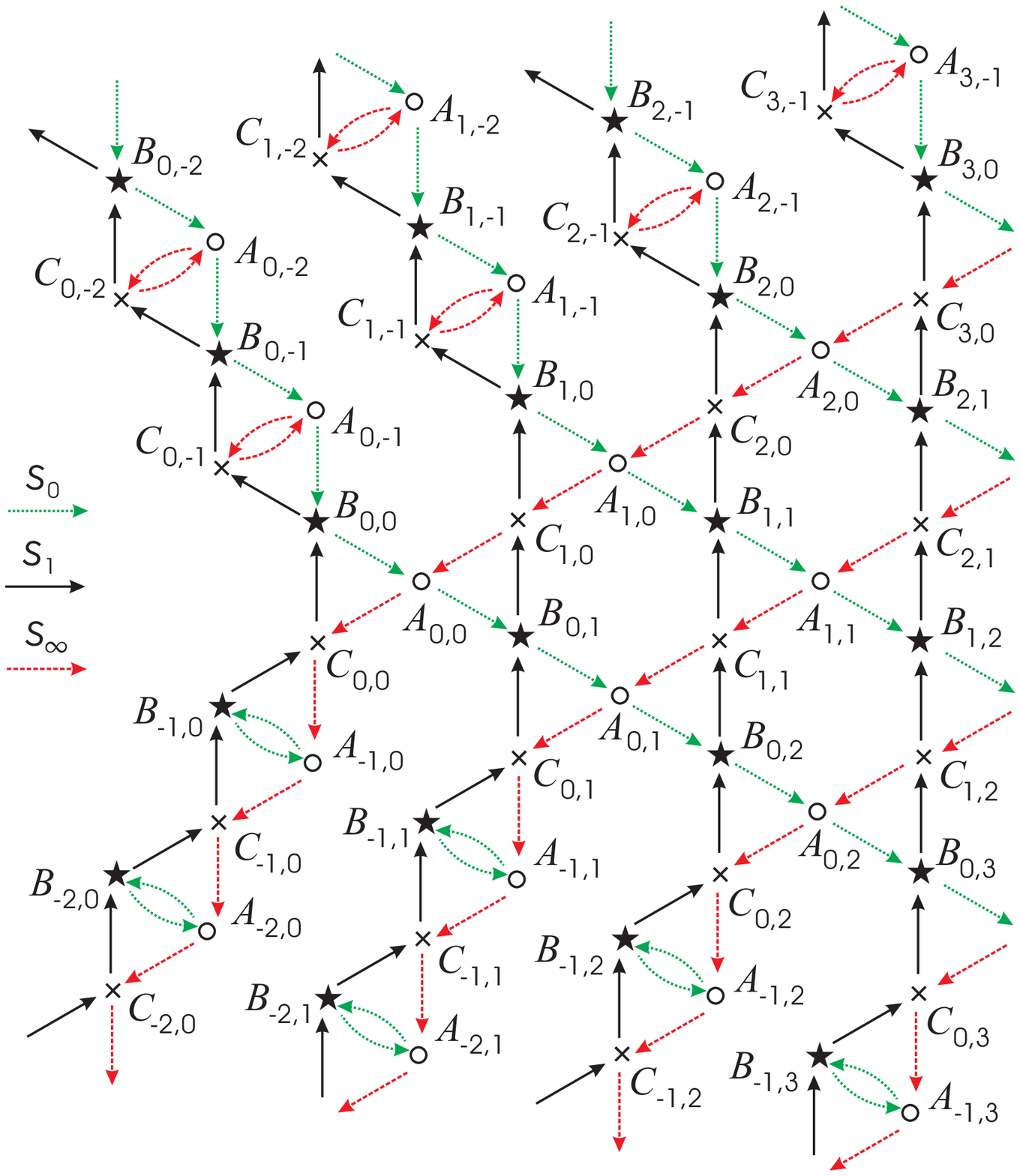}}
\nopagebreak
\vspace{.2in}

\begin{center}
\noindent Fig C11. Monodromy action for the cubic
\end{center}
\newpage

{\bf Quasi-Exactly Solvable quartics} (Bender-Boettcher)
\vspace{.1in}

The problem
\begin{equation}\label{8}
-y''+(-z^4-2\alpha z^2-2imz)\,y=\lambda z,\quad m\ge 1
\end{equation}
\begin{equation}\label{9}
y(r\,e^{i\theta})\to 0\;\mbox{as}\;r\to\infty,\quad\theta\in\{-\pi/6,-\pi+\pi/6\}
\end{equation}
is $PT$-symmetric for real $\alpha$.

There are $m$ ``elementary'' eigenfunctions
$$P(z)\exp(-iz^3/3-ibz)$$
$P$ polynomial of degree $m-1$.

Let $Z_m$ be the set of all pairs $(\alpha,\lambda)$ such
that $\lambda$ is an eigenvalue of (8), (9) corresponding to
an elementarey eigenfunction.
\vspace{.2in}

\noindent
{\bf Theorem 4} {\em For each $m$, the set $Z_m$ is irreducible.}

\vspace{.2in}
There are six Stokes sectors, with $S_0=\{-\pi/3<\arg z<0\}$
and the asymptotic values can be placed at the
points $(0,-1,0,1,0,\infty)$, listed anticlockwise,
starting from $S_0$.
The cell decomposition of the Riemann sphere is the same
as in Fig. C1, and its preimage is a line complex.
The associated trees have five ends, and the faces
labeled with $0$ have disjoint boundaries. Such trees
are classified into types A, B and C.
The deformation paths are the same as for the
PT-symmetric cubic, Fig. C10.  
\newpage
\begin{center}
\epsfxsize=4.0in%
\centerline{\epsffile{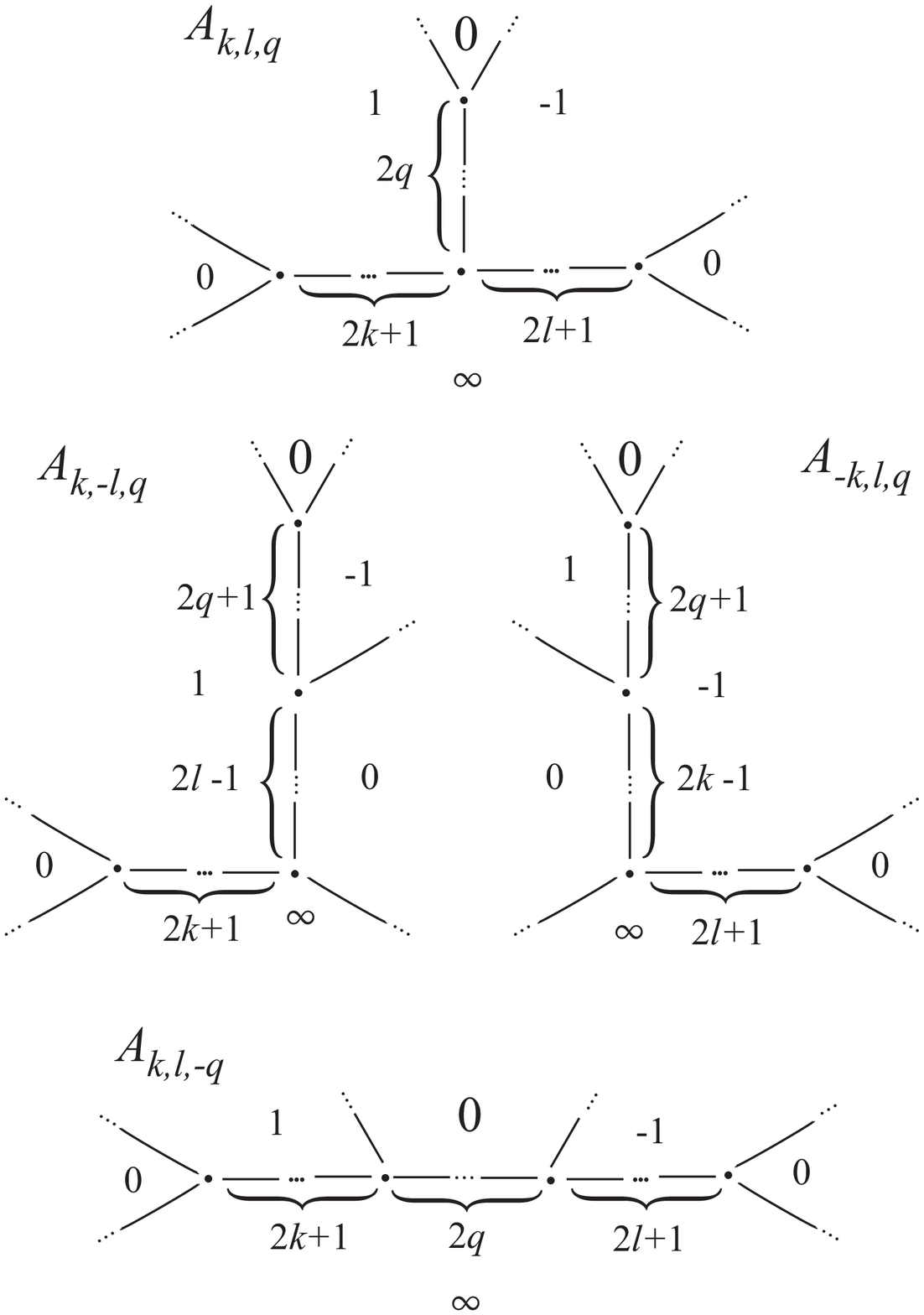}}
\nopagebreak
\vspace{.2in}

\noindent Fig Q1. Trees for the QES quartic, type $A$
\epsfxsize=4.0in%
\centerline{\epsffile{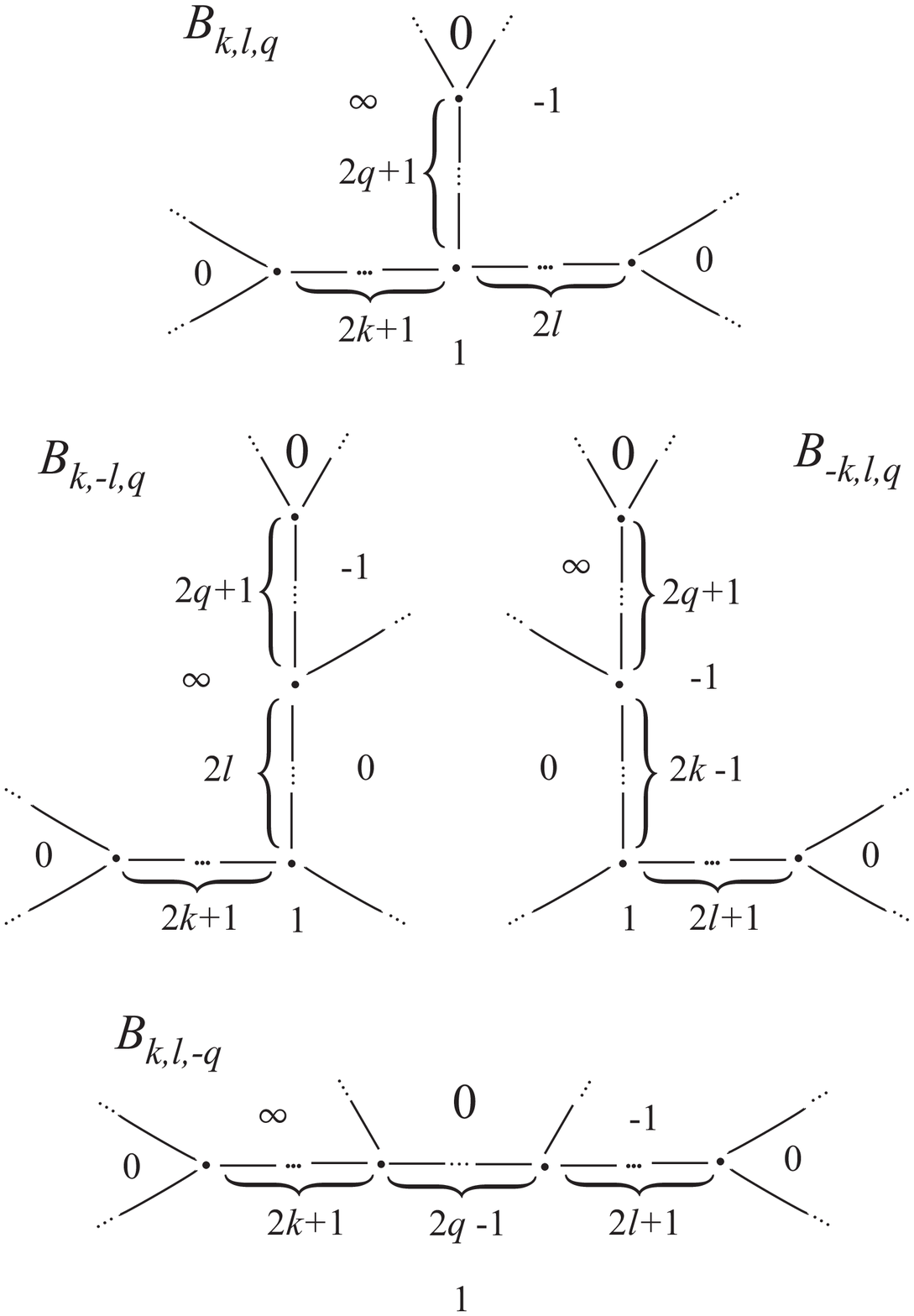}}
\nopagebreak
\vspace{.2in}

\noindent Fig Q2. Trees for the QES quartic, type $B$
\newpage

\epsfxsize=4.0in%
\centerline{\epsffile{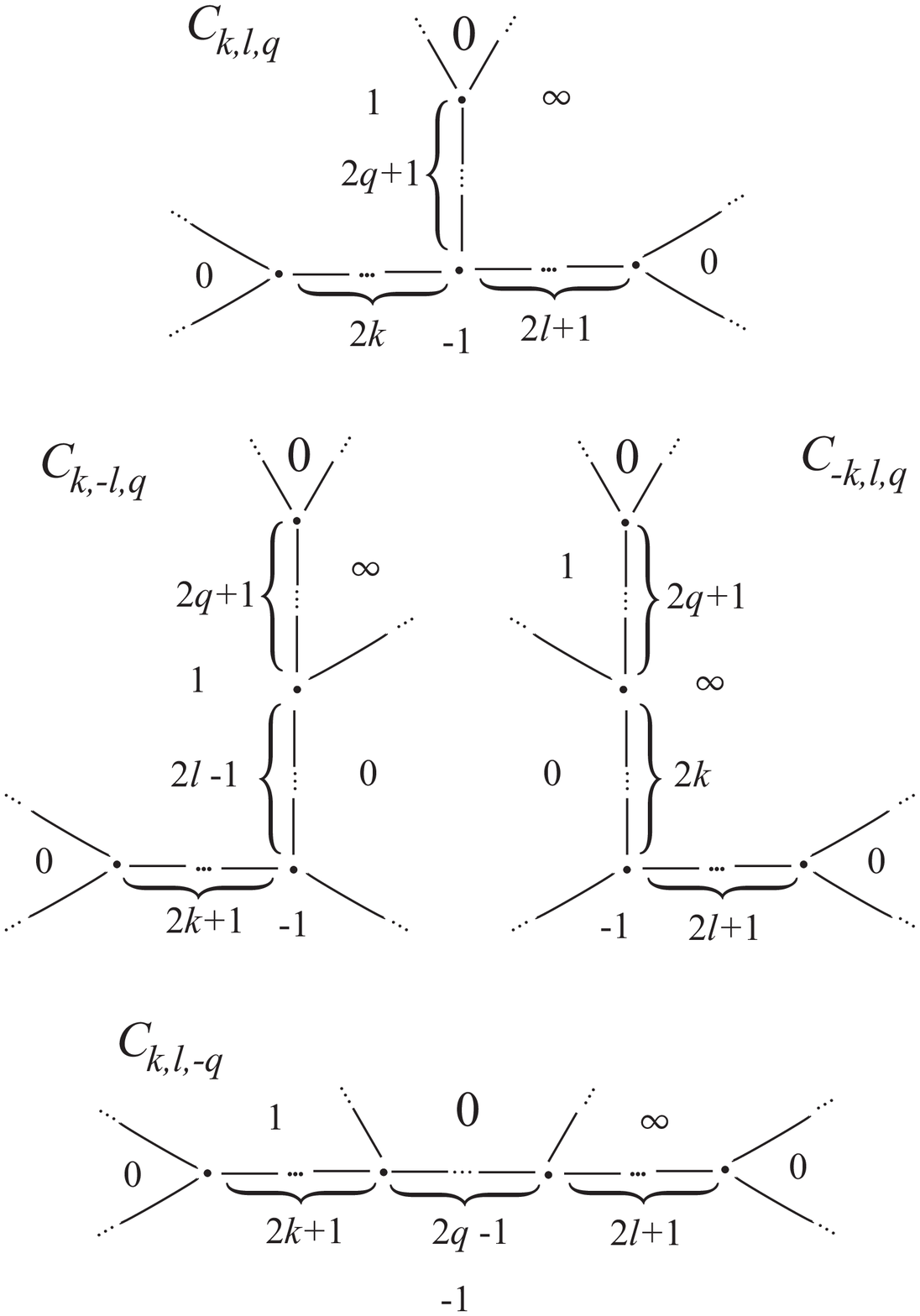}}
\nopagebreak
\vspace{.2in}

\noindent Fig Q3. Trees for the QES quartic, type $C$
\newpage
\vspace{-0.4in}

\epsfxsize=6.0in%
\centerline{\epsffile{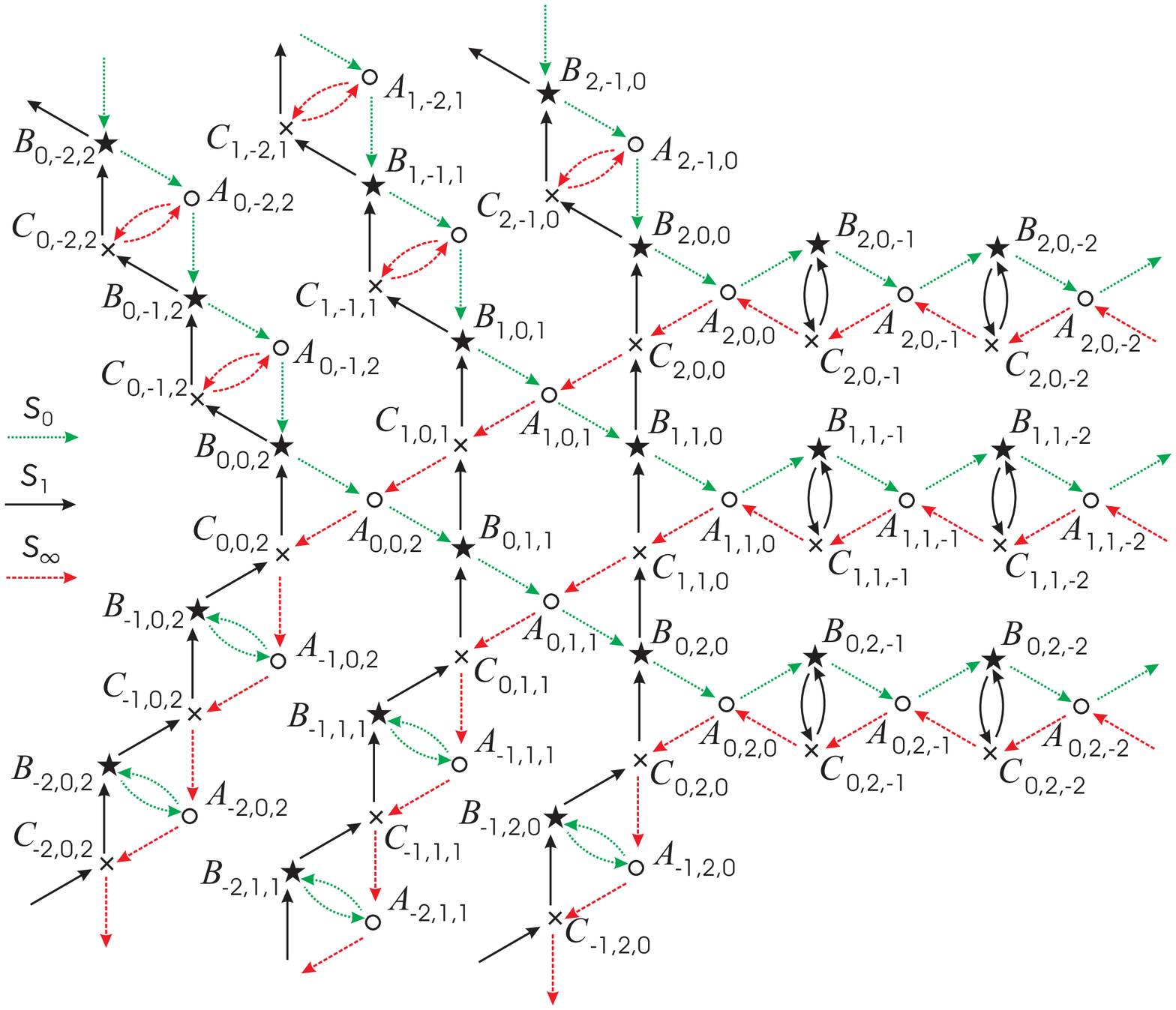}}
\nopagebreak
\vspace{.2in}

\noindent Fig Q4. Monodromy action for the QES quartic, $m=2$
\end{center}

www.math.purdue.edu/ $\tilde{}$ eremenko

www.math.purdue.edu/ $\tilde{}$ agabriel

\end{document}